\documentclass[aps,prf,a4paper,groupedaddress]{revtex4-2}
\usepackage[utf8]{inputenc} 

\usepackage{dcolumn}
\usepackage{bm}
\usepackage{amsfonts}
\usepackage{amsmath}
\usepackage{amssymb}
\usepackage{amsbsy}
\usepackage{graphicx}
\usepackage{psfrag}
\usepackage{subfigure}
\usepackage{verbatim}
\usepackage{textcomp}
\usepackage{color}
\usepackage{epsfig,bm,epsf,float}
\usepackage{mathrsfs}
\usepackage{wasysym}
\usepackage{comment}
\usepackage{txfonts}

\catcode`\@=11
\def\gtsim{\mathrel{\vcenter{\m@th\offinterlineskip
\hbox{$\hfill>\hfill$}\kern.5ex\hbox{$\hfill\sim\hfill$}}}}
\catcode`\@=12

\catcode`\@=11
\def\ltsim{\mathrel{\vcenter{\m@th\offinterlineskip
\hbox{$\hfill<\hfill$}\kern.5ex\hbox{$\hfill\sim\hfill$}}}}
\catcode`\@=12 \catcode`\@=12

\newcommand{\Rey}{\text{Re}}
\newcommand{\Sry}{\text{Sr}}
\newcommand{\Rry}{\text{Rr}}

\DeclareMathOperator{\sgn}{sgn}

\begin{document}
\preprint{Submitted to Physical Review Fluids}

\title{Drag and lift forces on a rigid sphere immersed in a wall-bounded linear shear flow}
\author{Pengyu Shi$^{1,2}$}
\author{Roland Rzehak$^1$}
\author{Dirk Lucas$^1$}
\author{Jacques Magnaudet$^{3,}\,$\footnote{{Email address for correspondence: jmagnaud@imft.fr}}}

\affiliation{$^1$Helmholtz-Zentrum Dresden – Rossendorf, Institute of Fluid Dynamics, Bautzner Landstrasse 400, D-01328 Dresden, Germany}
 \affiliation{$^2$Technische Universität Dresden, Faculty of Mechanical Engineering, Institute of Power Engineering, D-01062 Dresden, Germany}
 \affiliation{$^3$Institut de M\'ecanique des Fluides de Toulouse (IMFT), Universit\'e de Toulouse, CNRS, Toulouse, France}


\begin{abstract}

We report on a series of fully resolved simulations of the flow around a rigid sphere translating steadily near a wall, either in a fluid at rest or in the presence of a uniform shear. Non-rotating and freely rotating spheres subject to a torque-free condition are both considered to evaluate the importance of spin-induced effects. The separation distance between the sphere and wall is varied from values at which the wall influence is weak down to gaps of half the sphere radius. The Reynolds number based on the sphere diameter and relative velocity with respect to the ambient fluid spans the range $0.1-250$, and the relative shear rate defined as the ratio of the shear-induced velocity variation across the sphere to the relative velocity is varied from $-0.5$ to $+0.5$, so that the sphere either leads the fluid or lags behind it. The wall-induced interaction mechanisms at play in the various flow regimes are analyzed qualitatively by examining the flow structure, especially the spanwise and streamwise vorticity distributions.  Variations of the drag and lift forces at low-but-finite and moderate Reynolds number are compared with available analytical and semiempirical expressions, respectively. In more inertial regimes, empirical expressions for the two force components are derived based on the numerical data, yielding accurate fits valid over a wide range of Reynolds number and wall-sphere separations for both non-rotating and torque-free spheres.

\end{abstract}

\maketitle
\section{Introduction}\label{sec:1}

Determining the forces acting on particles moving parallel to a wall in a shear flow is of primary importance to understand and predict many features of wall-bounded particle-laden flows. In particular, the wall-normal force component governs crucial phenomena characterizing the dynamics and transfer properties in these flows, such as particle deposition, resuspension, saltation and near-wall preferential concentration. This force, albeit usually small in magnitude, plays a central role in separation techniques involving nearly neutrally buoyant particles, such as field-flow fractionation or crossflow filtration. Considering very dilute suspensions in which inter-particle or wall-particle collisions and direct hydrodynamic interactions play little role, quantitative predictions of how the particles move within the fluid and how in turn their presence possibly affects the flow require accurate expressions for the forces acting on an isolated particle to be available. The present work aims at contributing to this goal by considering a variety of near-wall configurations and flow regimes, identifying the dominant physical mechanisms at play in each of them, and providing accurate fits for the drag and lift components of the force acting on a spherical particle translating with respect to the wall and obeying either a non-rotating or a torque-free condition.

Due to its symmetrical shape and to the reversibility of Stokes equations, a sphere does not experience any lift force in the creeping-flow regime \citep{1962_Bretherton}. Therefore, this force arises through inertial effects associated with the ambient shear and/or the sphere translation and/or rotation with respect to the ambient flow.  
In a fluid with kinematic viscosity $\nu$, inertial effects associated with these three contributions become comparable to viscous effects at a distance $r$ from the sphere center such that
\begin{equation}
r\sim\textit{O} (\widetilde L_u)\,, \quad
r\sim\textit{O} (\widetilde L_\omega)\,, \quad
r\sim\textit{O} (\widetilde L_\Omega)\,,
\label{eq:inert_length}
\end{equation}
respectively. 
In \eqref{eq:inert_length}, $\widetilde L_u=\nu/|U_\text{rel}|$, $\widetilde L_\omega=(\nu/\gamma)^{1/2}$ and $\widetilde L_\Omega=(\nu/\Omega)^{1/2}$ are the so-called Oseen, Saffman and Magnus lengths, respectively, $U_\text{rel}$, $\gamma$ and $\Omega$ denoting the relative (or slip) velocity between the sphere and fluid, the shear rate in the undisturbed flow, and the norm of the sphere rotation rate $\boldsymbol{\Omega}$, respectively. In an unbounded shear flow, the vorticity generated at the sphere surface is advected asymmetrically in the wake by the ambient shear, yielding a transverse pressure gradient at distances of $\mathcal{O}(\widetilde L_\omega)$ downstream from the sphere, which results in a lift force directed toward the high- (low-) velocity side if the sphere lags behind (leads) the fluid. A similar mechanism is involved at distances of $\mathcal{O}(\widetilde L_\Omega)$ in the wake of a spinning sphere translating in a fluid at rest, and results in a Magnus or spin-induced lift force. A closed-form expression was obtained for this force in \cite{1961_Rubinow}, assuming  the slip and rotation Reynolds numbers to be small. If the sphere obeys a torque-free condition, as freely moving particles usually do if they do not collide with another particle or a wall, the spinning rate remains slow, implying $ \widetilde L_\Omega > \max(\widetilde L_u,\widetilde L_\omega) $. In this case, 
the parameter $ \varepsilon=\widetilde L_u/\widetilde L_\omega$ determines whether inertial effects are rather dominated by the ambient shear ($\varepsilon>1$) or the particle slip ($\varepsilon<1$). Saffman \citep{1965_Saffman, 1968_Saffman} considered a small sphere  translating in an unbounded linear shear flow and obtained the shear-induced lift force in closed form in the limit $ \varepsilon \gg1 $, assuming the slip and shear Reynolds numbers based on the sphere radius to be small. His prediction was extended to finite $ \varepsilon $ in \cite{1989_Asmolov} and \cite{1991_McLaughlin}, the results revealing that the lift force strongly decreases as the relative influence of the sphere translation increases, i.e. as $ \varepsilon$ decreases. Experiments \cite{1994b_Cherukat} and simulations \cite{1999_Cherukat} have confirmed these predictions down to $\varepsilon\approx0.4$ for particles with slip Reynolds numbers up to unity. Further insight into the shear-induced lift force in an unbounded fluid at higher Reynolds number was obtained through numerical studies \cite{1999_Kurose,2002_Bagchi_2}, revealing in particular that the distribution of the viscous stresses in the recirculating flow region at the back of the sphere makes this force reverse beyond a slip Reynolds number of some tens. Influence of the sphere rotation on the drag and lift forces in inertia-dominated regimes has also been examined, both for an imposed spinning motion and a torque-free condition \cite{1999_Kurose,2002_Bagchi}.\\
\indent When the flow is bounded by a single flat wall, the separation distance $ \widetilde L $ from the sphere center to the wall competes with the above three visco-inertial length scales through the ratios 
\begin{equation}
 L_u=\widetilde L / \widetilde L_u=\frac{\widetilde L|U_\text{rel}| }{\nu}\,,\quad L_\omega=\widetilde L / \widetilde L_\omega=\frac{\widetilde L\gamma^{1/2}}{\nu^{1/2}}\,,\quad L_\Omega=\widetilde L / \widetilde L_\Omega=\frac{\widetilde L\Omega^{1/2}}{\nu^{1/2}}\,,
\end{equation}
 which may be thought of as Reynolds numbers based on $ \widetilde L $ rather than on the particle size. 
In the sense of matched asymptotic expansions, the wall is located in the inner region of the disturbance if $ \max(L_u,L_\omega,L_\Omega)<1 $ (strictly speaking $\ll1$), while it stands in the outer region otherwise. Fundamental results were established by Cox \& Brenner \cite{1968_Cox} in the former case, showing in particular that, owing to the screening effect exerted by the wall, the leading-order estimate of the lift force may then be obtained through a regular expansion procedure. This work also enlightened the manner in which the generalized reciprocal theorem may be employed to obtain the lift force in the form of a volume integral solely involving creeping-flow solutions past the sphere. 

Asymptotic predictions for the slip-induced lift force acting on rigid spheres sedimenting close to a vertical wall in a fluid at rest in the low-Reynolds-number regime were obtained in \cite{1977_Cox} (based on the results of \cite{1968_Cox}) and \cite{1977_Vasseur}, assuming that the wall lies in the inner and outer regions of the disturbance, respectively. 
In this configuration, the lift force always tend to repel the particle from the wall and decreases gradually with increasing $L_u$. Considering the physical origin of the wall-particle interaction responsible for this force, which directly stems from wall-induced corrections to the flow in the wake region, the prediction of \cite{1977_Vasseur} was extended in a semiempirical manner up to slip Reynolds numbers of $\mathcal{O}(100)$, based on experiments performed with contaminated nearly-spherical air bubbles \citep{2003_Takemura}. Fully resolved simulations \citep{2005_Zeng,2009_Zeng} subsequently confirmed this semiempirical prediction.



Still in the low-but-finite Reynolds number regime, predictions for the shear-induced lift force in the presence of a wall standing in the inner region of the disturbance were also obtained in \cite{1977_Cox}, both for neutrally buoyant and negatively or positively buoyant particles. These results were then extended to the case of a wall standing in the outer region, first in the limit $\varepsilon\gg1$ \cite{1989_Asmolov}, then for arbitrary $\varepsilon$ \cite{1990_Asmolov,1993_McLaughlin}. These predictions were found to be valid up to slip Reynolds numbers of order unity in experiments performed under conditions $\varepsilon\lesssim1$ \cite{2009_Takemura_a,2009_Takemura_b}. They bridge the gap  between those of \cite{1977_Cox} and \cite{1991_McLaughlin} (hence \cite{1965_Saffman} in the limit $\varepsilon\gg1$), the latter being recovered in the limit where the wall is moved to infinity. While the slip-induced and shear-induced contributions to the lift superimpose linearly when the wall stands in the inner region, they are intrinsically coupled otherwise, owing to the nonlinear nature of the Oseen equation. Both contributions are directed away from the wall if the sphere lags behind the fluid, which is the case for a negatively (positively) buoyant particle in an upward (downward) shear flow near a vertical wall. Conversely, if the sphere leads the fluid, as for a light (heavy) particle in an upward (downward) shear flow, the shear-induced contribution tends to attract it toward the wall. In this case, the total lift force is attractive for large enough separations, but becomes repulsive again for short separations. This is because the wall influence gradually weakens the shear-induced contribution as the separation decreases, making the slip effect eventually dominant very close to the wall. In the above studies, the wall was considered sufficiently distant from the particle for the latter to be shrunk to a point. Obviously, this approximation is not tenable when the separation becomes of the order of a few sphere radii or less. Higher-order corrections accounting for the sphere finite size were obtained through the `reflection' technique \cite{2003_Magnaudet} (see also appendix A of \cite{1994_Cherukat}), but this approach cannot deal with situations in with the gap is less than typically the sphere radius. The combined use of exact creeping-flow solutions based on bi-spherical coordinates and the generalized reciprocal theorem allowed rational fits for the various contributions to the lift force to be obtained down to very small separations for both non-rotating and torque-free spheres \cite{1994_Cherukat, 1995_Cherukat,2010_Yahiaoui}. The limit case of a sphere held fixed on the wall, and that of a freely sliding and rolling sphere were worked out in \cite{1985_Leighton} and \cite{1995_Krishnan}, respectively.\\
\indent Numerical studies of hydrodynamic forces in near-wall configurations are quite scarce, presumably because they demand accurate boundary-fitted grids or refined immersed boundary techniques to properly capture the flow within the wall-particle gap. Variations of the slip- and shear-induced drag and lift forces in the low-but-finite Reynolds number regime with the wall located in either region of the disturbance were recently explored for both neutrally buoyant particles \cite{2020_Ekanayake_a} and arbitrarily buoyant particles \cite{2020_Ekanayake_b}. The characteristics of the slip-induced lift force in a wall-bounded fluid at rest were examined in detail in \cite{2005_Zeng} and \cite{2009_Zeng} down to small gaps and from Reynolds numbers of $\mathcal{O}(1)$ up to a few hundred. The same range of separations and Reynolds numbers was considered in \cite{2009_Zeng}  for a sphere held fixed with respect to the wall in a linear shear flow, a very specific choice corresponding to $L_u=L_\omega^2$. Near-wall forces acting on a sphere forced to spin in a fluid at rest were determined in \cite{2010_Lee} over a quite similar range of parameters, together with those experienced by a sphere immersed in a shear flow which either slides on the wall or spins very close to it.\\
\indent From the above review it appears that no study has considered inertia-dominated regimes for an arbitrarily translating and possibly freely rotating rigid sphere immersed in a wall-bounded shear flow, a situation of particular relevance to the widely encountered case of buoyant particles moving near a vertical wall. This is the problem addressed in the present work. The same problem was recently considered in \cite{2020_Shi_a} for spherical bubbles with a clean, i.e. shear-free, surface. Compared to the rigid sphere case, this difference in the dynamic boundary condition at the particle surface is known to affect the magnitude of the wall-induced forces in the low-Reynolds-number regime, but not the manner they vary with the flow parameters \cite{2003_Magnaudet}. This is no longer the case beyond Reynolds numbers of a few units, due to the much larger amount of vorticity produced at the sphere surface when the no-slip condition applies. In particular this difference results in the fact that the flow does not separate past a spherical bubble even at large Reynolds number while it does past a rigid sphere beyond a Reynolds number of $\mathcal{O}(10)$. 

In what follows we report on the results of fully resolved simulations of the flow past a freely translating and possibly rotating sphere. The sphere is immersed in a wall-bounded linear shear flow, may either lead the fluid or lag behind it, and obeys a non-rotating or a torque-free condition. 
In Sec.  \ref{sec:2} we formulate the problem, specify the considered range of parameters and outline the numerical approach which is essentially similar to that employed in \cite{2020_Shi_a}. Section \ref{sec:3} summarizes theoretical and semiempirical expressions for the forces acting on a sphere in an unbounded shear flow and in wall-bounded configurations. Numerical results are first used in Sec. \ref{sec:4} to examine the physical mechanisms induced by the presence of the wall and the corresponding alterations of the near-sphere flow in the various regimes. Variations of the drag and lift forces with the flow parameters are analyzed in Sec. \ref{sec:5}. Empirical fits reproducing the observed variations in specific regimes or throughout the entire parameter range of the simulations are established. A summary of the main outcomes, especially regarding these empirical fits of direct interest in applications, is provided in Sec. \ref{sec:6}.

\section{Statement of the problem and outline of the simulation approach}\label{sec:2}

\begin{figure}
  \centerline{\includegraphics[scale=1]{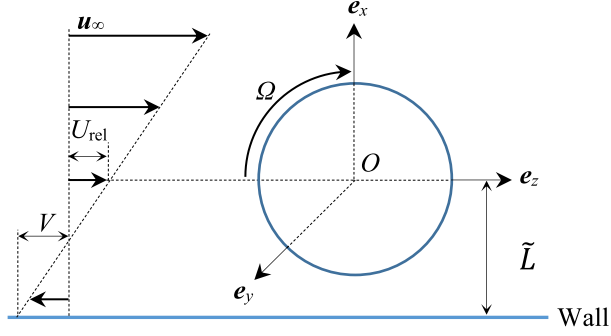}}
  \caption{Schematic of a sphere moving in a wall-bounded linear shear flow.}
\label{fig:schem_bub_mov}
\end{figure}

We define a Cartesian coordinate system $(Oxyz)$ with the origin located at the center of the sphere, as illustrated in Fig. \ref{fig:schem_bub_mov}. We assume that the sphere moves parallel to a single planar wall with a translational velocity $\boldsymbol{V}=V\boldsymbol{e}_z$ and a rotational velocity $\boldsymbol{\Omega}=-\Omega\boldsymbol{e}_y$. The wall is located at $x=-\widetilde L$ and $\boldsymbol{e}_x$ denotes the wall-normal unit vector pointing into the fluid. In the reference frame translating with the sphere, the undisturbed flow is a one-dimensional linear shear flow with a velocity profile $\boldsymbol{u}_\infty=[\gamma(\widetilde L+x)-V]\boldsymbol{e}_z$ and a spanwise vorticity $\boldsymbol{\omega}_\infty=-\gamma \boldsymbol{e}_ {y}$. The relative (or slip) velocity of the fluid with respect to the sphere is then $\boldsymbol{U}_\text {rel}=(\gamma\widetilde L-V)\boldsymbol{e}_z$. The fluid velocity and pressure fields in the presence of the sphere are denoted by $\boldsymbol{u}$ and $p$, respectively, and $\boldsymbol\omega={\nabla}\times\boldsymbol{u}$ denotes the vorticity. 

Assuming the fluid to be Newtonian and considering the flow as incompressible, the continuity and Navier-Stokes equations read 
\refstepcounter{equation}
$$
\nabla\cdot \boldsymbol{u}=0\,; \quad
{\frac{\partial \boldsymbol{u}}{\partial t}+\boldsymbol{u}\cdot  \nabla\boldsymbol{u}=-\frac{1}{\rho } \nabla p+ \nu\nabla^2\boldsymbol{u}}\,,
\eqno{(\theequation{\text{a},\text{b}})}
\label{eq:ns}
$$
with $\rho$ and $\nu$ the fluid density and kinematic viscosity, respectively. Boundary conditions at the sphere surface, at the wall, and in the far field read, respectively
\begin{equation}
\boldsymbol{u}=\left\{
\begin{array}{ll}
\boldsymbol{\Omega} \times \boldsymbol{r} & \mbox{for}\quad r=d/2\,, \\[2pt]
-V\boldsymbol{e}_z & \mbox{for}\quad x=-\widetilde L\,, \\[2pt]
\boldsymbol{u}_\infty=[\gamma (\widetilde L+x)-V]\boldsymbol{e}_z &  \mbox{for}\quad r\to \infty\,,
\end{array} \right.
\label{eq:BC1}
\end{equation}
where  $r=(x^2+y^2+z^2)^{1/2}$ denotes the distance to the sphere center, and $d$ is the sphere diameter. 

With the above boundary conditions, the steady flow field past the sphere depends on four characteristic parameters, namely the slip Reynolds number, $\Rey$, the dimensionless shear rate, $\Sry$, the dimensionless sphere rotation rate, $\Rry$, and the normalized wall distance, $L_\text{R}$. These control parameters are defined as 
\begin{equation}
\Rey=\frac{|U_\text {rel}|d}{\nu}, \quad
\Sry=\frac{\gamma d}{U_\text {rel}}, \quad
\Rry=\frac{\Omega d}{U_\text {rel}}, \quad
L_\text{R}=\frac{2\widetilde L}{d},
\label{eq:def_var}
\end{equation}
with $U_\text {rel}=\boldsymbol{U}_\text {rel}\cdot \boldsymbol{e}_ {z}$ and $\Omega=-\boldsymbol{\Omega}\cdot \boldsymbol{e}_ {y}$. Under the torque-free condition, $\Rry$ is entirely determined by the other three parameters and is no longer an independent control parameter. A positive (negative) $\Sry$ indicates that the sphere lags (leads) the fluid, the former case corresponding to the flow configuration sketched in Fig. \ref{fig:schem_bub_mov}. 
In most of this work, $\Rey$, $\Sry$, and $L_\text{R}$ are varied in the range $[0.1, 250]$, $ [-0.5, 0.5]$, and $[1.5, 8]$, respectively. Hence $\varepsilon$ is up to $2.2$ for $\Rey=0.1$ and becomes less than $1$ as soon as $\Rey>0.5$, and even less than $0.1$ beyond $\Rey=50$. 
In an unbounded fluid, the shear-induced transverse force is proportional to $\Sry$, so that its sign changes with that of $U_\text {rel}$. In the presence of a nearby wall, three different regimes are encountered. If $\Sry\ll1$, the transverse force results primarily from the particle relative translation with respect to the wall, which at low-but-finite Reynolds number makes it proportional to $\Rey$. In contrast, when $\Sry\gg1$, i.e. the slip velocity is small compared to the shear-induced velocity variation at the particle scale, the dominant contribution to the transverse force is proportional to $\Sry^2$. This regime, relevant to small nearly neutrally buoyant particles, will not be considered here (it was recently specifically examined in \cite{2018_Asmolov, 2020_Ekanayake_b} in the low-$\Rey$ range, with applications to inertial microfluidics in mind). 
The near-wall transverse force does not change sign with  $U_\text {rel}$ in the above two regimes, being repulsive in both cases. In contrast, it may change sign  when $\Sry\lesssim1$, which is the regime we are primarily interested in.\\
\indent Let us briefly illustrate some flow configurations covered by the above parameter range. Consider for instance a $1\,$mm-diameter particle sedimenting in water and assume the particle is twice as dense as the fluid. Then the standard drag law predicts that its slip Reynolds number is approximately $115$. With $|\Sry|=0.5$, this yields $\gamma\approx57\,\text{s}^{-1}$. This is for instance the near-wall shear rate in the laminar flow in a $15\,$mm-high plane channel, the corresponding Reynolds number $\text{Re}_\text{H}$ based on the depth-averaged velocity being $\text{Re}_\text{H}\approx2200$. In the same configuration, a particle ten times smaller ($d=0.1\,$mm) has a slip Reynolds number close to $0.55$ and the shear rate corresponding to $|\Sry|=0.5$ is $27\,\text{s}^{-1}$, the near-wall value reached in a $1.5\,$mm-high channel with $\text{Re}_\text{H}\approx10$. Consider now that the largest of the above two particles is immersed in a vertical turbulent boundary layer and stands $1\,$mm apart from the wall (which corresponds to $L_\text{R}=2$) in the logarithmic region. Then, equating $\gamma$ to the local time-averaged shear rate $u^*/(\kappa L_\text{R})$, with $\kappa=0.4$ the von K\'arm\'an constant, the associated friction velocity $u^*$ is close to $2.25\,\text{cm.s}^{-1}$, which corresponds to an outer velocity close to $0.6\,\text{m.s}^{-1}$, i.e. $\text{Re}_\text{H}\approx3\times10^4$ if the flow takes place in a  $5\,$cm-high channel. Still with $L_\text{R}=2$, the $d=0.1\,$mm particle rather stands within the viscous sublayer. There, the time-averaged shear rate corresponding to $|\Sry|=0.5$ is $\gamma=u*^2/\nu\approx27\,\text{s}^{-1}$, which yields $u^*\approx5.2\,\text{mm.s}^{-1}$, hence an outer velocity close to $15\,\text{cm.s}^{-1}$ corresponding to $\text{Re}_\text{H}\approx7500$. 

\indent In the present problem, the drag force parallel to $\boldsymbol{U}_\text {rel}$, $F_\text D$, the lift force parallel to $\boldsymbol{e}_x$, $F_\text L$, and the torque antiparallel to $\boldsymbol{e}_y$, $M$, acting on the sphere are defined as
\begin{equation}
F_\text D=\frac{\boldsymbol{U}_\text {rel}}{||\boldsymbol{U}_\text {rel}||}\cdot \int_\Gamma \boldsymbol{\Sigma} \cdot \boldsymbol{n}\:\mathrm{d} \Gamma, \quad
F_\text L=\boldsymbol{e}_x \cdot \int_\Gamma \boldsymbol{\Sigma} \cdot \boldsymbol{n}\:\mathrm{d} \Gamma, \quad
M=-\boldsymbol{e}_y \cdot \int_\Gamma \boldsymbol{r}\times (\boldsymbol{\Sigma} \cdot \boldsymbol{n}) \:\mathrm{d} \Gamma, 
\label{eq:def_force}
\end{equation}
where $\boldsymbol{\Sigma}$ is the stress tensor and $\boldsymbol{n}$ denotes the outward unit normal to the sphere surface $\Gamma$. Results concerning the two force components will be expressed in terms of the lift and drag coefficients, $C_\text L$ and $C_\text D$, obtained by dividing the corresponding force by $\pi d^2\rho U_\text {rel}^2/8$. According to the above definition, a negative (positive) $C_\text L$ corresponds to a force directed toward (away from) the wall. In the case of a non-rotating sphere, results concerning the hydrodynamic torque will be expressed using the torque coefficient $C_\text M=M/(\pi d^3\rho U_\text {rel}^2/16)$. We use the notations $C_\text D^\text W$ ($C_\text L^\text W$) and $C_\text D^\text U$ ($C_\text L^\text U$) to denote the drag (lift) coefficients in wall-bounded and unbounded flows, respectively. Situations where the wall lies in the inner or outer region of the disturbance will be distinguished by superscripts $\text{W-in}$ and $\text{W-out}$, respectively. Results for the drag coefficient are usually given in the form of the relative wall-induced change $\Delta C_\text D=(C_\text D^\text W-C_{\text D0}^\text U)/C_{\text D0}^\text U$, with $C_{\text D0}^\text U$ denoting the drag coefficient on a sphere translating in an unbounded uniform fluid. Drag (lift) contributions corresponding to the slip-induced effect are denoted with the subscript $\text Du$ ($\text Lu$), while those corresponding to the shear-induced effect are denoted with the subscript $\text D\omega$ ($\text L\omega$). Similar conventions are applied to the rotation rate, $\Rry$.

The three-dimensional flow field past the sphere is computed with the JADIM code developed at IMFT. The sphere center stands on the axis of a large cylindrical computational domain, one base of which coincides with the wall. The reader is referred to \cite{2020_Shi_a} for numerical aspects concerning the specificities of the code, the size of the computational domain (varied with $\Rey$ so as to minimize confinement effects in all flow regimes), the grid system and number of nodes in each direction (which also varies with $\Rey$, to ensure that the boundary layer and the near wake are properly resolved), and the boundary conditions. Comparison with available data and asymptotic predictions may be found in the same reference, together with grid convergence tests. To achieve the torque-free condition, the sphere rotation rate $\Rry$ is computed through an iterative approach. First, the steady flow around a non-rotating sphere is determined to obtain the corresponding torque coefficient, $C_\text {M,0}$. Then the rotation rate is updated as $\Rry_1 = \Rry_0+C_\text {M,0}\Rey/16$ and the corresponding steady flow field is determined to obtain the new torque coefficient,  $C_\text {M,1}$. This procedure is continued until the torque coefficient becomes less than $0.05C_\text {M,0}$. 

\section{Analytical solutions and empirical predictions}\label{sec:3}


\subsubsection{Unbounded linear shear flow}\label{sec:3.1}
\indent At low-but-finite Reynolds number, the presence of a uniform shear in an unbounded flow domain results in a transverse or lift force on a sphere in the direction of $\boldsymbol{U}_\text {rel}\times\boldsymbol{\omega}$. For $\Sry\gg1$, the leading-order force is proportional to $(|\Sry|/\Rey)^{1/2}$. In the case of a non-rotating sphere, the corresponding lift coefficient takes the form \citep{1965_Saffman, 1991_McLaughlin}
\refstepcounter{equation}
$$
C_{\text L\omega}^\text U(\Rey\ll1)=\frac{18}{\pi^2}\sgn(\Sry)\varepsilon J_\text L(\varepsilon) \quad\mbox{with}\quad
J_\text L(\varepsilon)\approx2.254(1+0.2\varepsilon^{-2})^{-3/2}\,,
\eqno{(\theequation{\text{a},\text{b}})}\label{eq:U_low_Re}
$$
where $\varepsilon=\widetilde L_u/\widetilde L_\omega=(|\Sry|/\Rey)^{1/2}$ is the ratio of the Oseen and Saffman lengths, and $\sgn(\Sry)=\Sry/|\Sry|$. 
Assuming $\Rey$ and $\Rey\Rry$ to be small, a rotating sphere translating in a fluid at rest experiences a transverse force $\frac{\pi}{8} \rho d^3 \boldsymbol{\Omega}\times\boldsymbol{U}_\text {rel}$ \citep{1961_Rubinow}, which yields a lift coefficient $C_{\text L\Omega}^\text U[\Rey\ll1]=\Rry$. When ambient shear and rotation act together, the total lift force including $\mathcal{O}(\Rry)$- and $\mathcal{O}(\Sry)$-effects is the sum of the two individual contributions, and involves a second-order shear-induced contribution lowering the lift coefficient by $-\frac{11}{8}\Sry$ in the limit $|\Sry|\gg1$ \citep{1965_Saffman}.
Therefore, the total lift coefficient  takes the form 
\begin{equation}
C_{\text L\omega}^\text U(\Rey\ll1)\approx\frac{18}{\pi^2}\sgn(\Sry)\varepsilon J_\text L(\varepsilon)-\frac{11}{8}\Sry+\Rry\,,
\label{eq:U_low_Re_tf}
\end{equation}
If the torque-free condition holds, the leading-order sphere rotation rate in the low-$\Rey$ regime is half the undisturbed flow vorticity, i.e. 
\begin{equation}
\Rry=\Rry^\text U(\Rey\ll1)\approx \frac{1}{2}\Sry \,.
\label{eq:spin_rate_u_lowRe}
\end{equation}
Consequently, the rotation-induced and second-order shear-induced lift forces combine in a correction of $-\frac{7}{8}\Sry$ to (\ref{eq:U_low_Re})  \citep{1965_Saffman}. \\
 At low-to-moderate Reynolds numbers, the shear-induced lift force predicted by \eqref{eq:U_low_Re} agrees well with numerical data for non-rotating spheres up $\Rey\approx10$ \citep{1999_Kurose}. Increasing $\Rey$, this force first exhibits weak positive values up to $\Rey\approx50$. Beyond this range, it changes sign, owing to the influence of the standing eddy on the stress distribution at the rear of the sphere. For $\Rey\gtrsim50$, the numerical results of \cite{1999_Kurose} and \cite{2002_Bagchi} are adequately fitted by the empirical correlation \cite{2008_Loth} 
 \begin{equation}
C_{\text L\omega}^\text U[\Rey=\mathcal{O}(100)]\approx -\sgn(\Sry)|\Sry|^{1/3}\left\{0.0525+0.0575\tanh\left[11.5\log\left(\frac{\Rey}{120}\right)\right]\right\}\,.
\label{eq:lift_loth}
\end{equation}
  The spin-induced lift coefficient $C_{\text L\Omega}^\text U(\Rey)$ remains linearly proportional to the rotation rate at moderate Reynolds number. Setting $C_{\text L\Omega}^\text U(\Rey)=c_{\text L\Omega}^\text U(\Rey)\Rry$, the coefficient $c_{\text L\Omega}^\text U$ is found to be smaller than unity and nearly independent of the Reynolds number for $\Rey\gtrsim1$.  According to \citep{2002_Bagchi}, one has at moderate $\Rey$ 
\begin{equation}
c_{\text L\Omega}^\text U[\Rey=\mathcal{O}(1-100)]\approx 0.55\,.
\label{eq:spin_lift_u}
\end{equation}
In the same range of Reynolds number, the torque-free spin rate normalized by the ambient rotation rate is found to depend only on $\Rey$ in the form \citep{2002_Bagchi}
\begin{equation}
\Rry=\Rry^\text U [\Rey=\mathcal{O}(1-100)]\approx  f_{\Omega}^\text U(\Rey)\frac{\Sry}{2}
\label{eq:spin_rate_ua}
\end{equation}
with
\begin{equation}
 f_{\Omega}^\text U(\Rey)\approx1-0.0364\Rey^{0.95}\quad\mbox{for}\quad 0.5\leq\Rey\leq5\quad\mbox{and}\quad f_{\Omega}^\text U(\Rey)\approx1-0.0755\Rey^{0.455}\quad \mbox{for}\quad\Rey>5 \,.
\label{eq:spin_rate_ub}
\end{equation}
Available DNS results for torque-free rotating spheres suggest that contributions of shear and rotation still superpose linearly in the lift force up to $\Rey=100$ \citep{2019_Shi}.

\subsubsection{Low-$Re$ wall-bounded shear flow}\label{sec:3.2}

The presence of a nearby wall results in a drag increase, while for reasons mentioned above it may either increase or decrease the transverse force, depending on the sign of $\Sry$. For $\Rey\ll1$, situations where the wall lies in the inner region of the disturbance,  i.e. $\max{(L_u, L_\omega, L_\Omega)} \ll1$, were investigated in \cite{1967_Goldman_a,1967_Goldman_b,1977_Cox,1994_Cherukat, 2003_Magnaudet}. 

In the case of a non-rotating sphere, the results of \cite{1994_Cherukat} indicate that the lift coefficient is approximately 
\footnote{
 In (\ref{eq:CLWin}), pre-factors expressed in fractional form were derived analytically by Lovalenti in an appendix to \cite{1994_Cherukat}, while those expressed in decimal form originate from the fitted value of the force computed in the form of a volume integral in the same reference.}
\begin{eqnarray}
\begin{aligned}
C_\text L^\text{W-in}(\Sry,L_\text{R})&= \underbrace{\frac{9}{8}\left(1+\frac{1}{8}L_\text{R}^{-1}-0.413L_\text{R}^{-2}+0.270L_\text{R}^{-3} \right)
	}_{C_{\text Lu}^\text{W-in}}
	 \\&+ \underbrace{
		\frac{33}{32}\left(L_\text{R}+\frac{17}{48}+0.643L_\text{R}^{-1}-0.280L_\text{R}^{-2} \right)\Sry+\frac{61}{192}\left(1+0.527L_\text{R}^{-1}-1.200L_\text{R}^{-2}+0.657L_\text{R}^{-3}\right)\Sry^2}_{C_{\text L\omega}^\text{W-in}}\,.
		\label{eq:CLWin}
	\end{aligned}
\end{eqnarray}
Similarly, in the case of a torque-free rotating sphere, one has
\begin{eqnarray}
\begin{aligned}
C_\text L^\text{W-in}(\Sry,L_\text{R})&= \underbrace{\frac{9}{8}\left(1+\frac{3}{16}L_\text{R}^{-1}-0.511L_\text{R}^{-2}+0.287L_\text{R}^{-3} \right)
	}_{C_{\text Lu}^\text{W-in}}
	 \\&+ \underbrace{
		\frac{33}{32}\left(L_\text{R}+\frac{443}{528}+0.258L_\text{R}^{-1}-0.145L_\text{R}^{-2}\right)\Sry+\frac{55}{192}\left(1+\frac{9}{16}L_\text{R}^{-1}-1.090L_\text{R}^{-2}+0.568L_\text{R}^{-3}\right)\Sry^2}_{C_{\text L\omega}^\text{W-in}}\,.
		\label{eq:CLWin_tf}
\end{aligned}
\end{eqnarray}
The difference between \eqref{eq:CLWin} and \eqref{eq:CLWin_tf} indicates an increase of the lift coefficient by $\frac{1}{2}\Sry$ and a decrease by $-\frac{1}{32}\Sry^2$ for large $L_\text{R}$, when switching from the zero-rotation condition to the zero-torque one. The $\frac{1}{2}\Sry$-increase is in line with the contribution of the torque-free rotation to the lift force found in the unbounded case.\\
\indent Still for a torque-free sphere, the dimensionless rotation rate is approximately \citep {1967_Goldman_a,1967_Goldman_b,2003_Magnaudet}
\begin{equation}
\Rry^\text{W-in}(\Sry,L_\text{R})\approx\underbrace{- \frac{3}{16}L_\text{R}^{-4}\left(1-\frac{3}{8}L_\text{R}^{-1}\right)
                                                                   }_{\Rry_u^\text{W-in}}
                                              +\underbrace{\frac{1}{2}\left(1-\frac{5}{16}L_\text{R}^{-3}\right)\Sry
                                                                   }_{\Rry_ \omega^\text{W-in}}\,,
\label{eq:spin_rate_w}
\end{equation}
while the variation of the drag force is \citep{2003_Magnaudet}
\begin{eqnarray}
\begin{aligned}
\Delta C_\text D^\text{W-in}(\Sry,L_\text{R})&= \underbrace{
		\left(   \frac{9}{16}L_\text{R}^{-1}-\frac{1}{8}L_\text{R}^{-3}+\frac{45}{256}L_\text{R}^{-4}+\frac{1}{16}L_\text{R}^{-5}  \right) 
		\left(    1-\frac{9}{16}L_\text{R}^{-1}+\frac{1}{8}L_\text{R}^{-3}-\frac{45}{256}L_\text{R}^{-4}-\frac{1}{16}L_\text{R}^{-5} \right)^{-1}
	}_{\Delta C_{\text Du}^\text{W-in}} 
	 \\& -  \underbrace{
		\frac{5}{32} \left(   L_\text{R}^{-2}+\frac{9}{16}L_\text{R}^{-3}                   \right)\Sry\,,
	}_{\Delta C_{\text D\omega}^\text{W-in}}\,
	\label{eq:CDWin_tf}
\end{aligned}
\end{eqnarray}
where $\Delta C_\text D^\text{W-in}(\Sry,L_\text{R})=\left(C_\text D^\text{W-in}(\Sry,L_\text{R})-C_\text {D0}^\text U(\Rey\rightarrow0)\right)/C_\text {D0}^\text U(\Rey\rightarrow0)$, with $C_\text {D0}^\text U(\Rey\rightarrow0)=24/\Rey$ the drag coefficient in the creeping flow limit. Since the leading contribution of the particle rotation to the drag force is known to be proportional to  $L_\text{R}^{-4}\Rry$ 
 \citep {1967_Goldman_a}, the above $\mathcal{O}(L_\text{R}^{-5})-\mathcal{O}(L_\text{R}^{-3}\Sry)$ approximation for $\Delta C_\text D^\text{W-in}(\Sry,L_\text{R})$ also holds for a non-rotating sphere.


When the wall lies in the outer region of the disturbance, the relative length scales $L_u$, $L_\omega$, and $L_\Omega$ are no longer small. Hence, in addition to $L_\text{R}$, the drag and lift forces depend on these three visco-inertial length scales. This situation was investigated in the shearless non-rotating case ($L_\omega \to 0$, $L_\Omega \to 0$) in \cite{1977_Vasseur}, neglecting the finite size of the particle. The relevant solutions were obtained in the form of double integrals which can be approximated as \citep{2020_Shi_a}
\begin{equation}
\frac{16}{9}L_\text{R}\Delta C_{\text Du}^\text{W-out}(\Rey\lesssim1)= f_\text D^{\prime}(L_u)\approx\frac{1}{1+0.16L_u(L_u+4)}\,,
\label{eq:fD_prime}
\end{equation}
and
\begin{equation}
\frac{8}{9} C_{\text Lu}^\text{W-out}(\Rey\lesssim1)=f_\text L^{\prime}(L_u)\approx\left\{
\begin{array}{ll}
[1+0.13L_u(L_u+0.53)]^{-1} & \mbox{for}\quad L_u\leq10\,, \\[2pt]
7.95L_u^{-2.09} &  \mbox{for}\quad L_u>10\,,
\end{array} \right.
\label{eq:fL_prime}
\end{equation}
with $\Delta C_{\text Du}^\text{W-out}(\Rey\ll1)=\left(C_{\text Du}^\text{W-out}(L_u,L_\text{R})-C_{\text D0}^\text U(\Rey\rightarrow0)\right)/C_{\text D0}^\text U(\Rey\rightarrow0)$. The two functions $f_\text D^{\prime}$ and $f_\text L^{\prime}$ describe how the wall-induced drag modification and the transverse force decay as inertial effects in the bulk become dominant compared to the wall influence. 

In the presence of shear, the case of a non-rotating sphere close to a wall standing in the outer region of the disturbance was worked out in \cite{1993_McLaughlin}. Again, the solution was obtained in the form of a volume integral in Fourier space. The value of this integral cannot be obtained in closed form but was tabulated for various values of $L_\omega$ and $ \varepsilon=\widetilde L_u/\widetilde L_\omega$. These results were fitted in \cite{2009_Takemura_a} to obtain tractable estimates of the lift force. This fit was further modified in \cite{2020_Shi_a} to take into account the effects of the finite particle size, which tend to lower the transverse force when the particle gets very close to the wall. The same argument was used to derive an empirical expression for the drag variation. Making use of the approximate expression (\ref{eq:U_low_Re}) for $C_{\text L\omega}^\text U(\Rey\ll1)$ and of the expression for $C_{\text Lu}^\text{W-in}$ in (\ref{eq:CLWin}), the final expression for the lift force in the case of a non-rotating sphere takes the form 
\begin{equation}
C_\text L^\text{W}(\Rey\lesssim1)\approx  
	 f_\text L(L_\omega,\varepsilon) f_\text L^{\prime}(L_u) C_{\text Lu}^\text{W-in}
	  + h_\text L(L_\omega,\varepsilon) C_{\text L\omega}^\text U(\Rey\ll1)\,,
\label{eq:CWlowRe}
\end{equation}
with $f'_\text{L}(L_u)$ as given in (\ref{eq:fL_prime}) and
\refstepcounter{equation}
$$
f_\text L(L_\omega,\varepsilon) = \exp{^{-0.22\varepsilon^{0.8} L_\omega^{2.5}}} \quad\mbox{and}\quad
h_\text L(L_\omega,\varepsilon) =1-\exp{^{-\frac{11}{96} \pi^2 \frac{ L_\omega}{J_\text L(\varepsilon)}
                                                              (1+\frac{17}{48}L_\text{R}^{-1}+0.643L_\text{R}^{-2}-0.280L_\text{R}^{-3})}}\,.
\eqno{(\theequation{\text{a},\text{b}})}\label{eq:fhL0}
$$
Thanks to these empirical pre-factors, (\ref{eq:CWlowRe}) approaches the inner solution (\ref{eq:CLWin}) when $L_u \to 0$ and $L_\omega \to 0$, with the exception of the $\Sry^2$-term, usually much smaller than the $\Sry$-term as far as $|\Sry|\lesssim1$. 
 Similarly, making use of (\ref{eq:CDWin_tf}) and (\ref{eq:fD_prime}), the total slip-induced near-wall correction to the drag taking into account the finite size of the sphere may be approximated as
\begin{equation}
\Delta C_\text D^\text{W}(\Rey\lesssim1)\approx  
f_\text D^{\prime}(L_u) \Delta C_{\text D}^\text{W-in}\,. \quad
\label{drag_fin}
\end{equation}

\subsubsection{Approximate expression for the slip-induced transverse force in a fluid at rest at moderate-to-large \text{Re}}\label{sec:3.3}	

No theoretical solution for the hydrodynamic loads can be found when inertial effects are dominant. However, reliable empirical extensions of the low-Reynolds-number predictions may be achieved based on accurate data. Several experimental and numerical studies \citep{2003_Takemura,2005_Zeng,2009_Zeng} examined the motion of a rigid sphere close to a wall in a quiescent fluid. They revealed that the transverse force exhibits a faster decay with increasing $L_u$ than predicted by the low-but-finite $\Rey$ solution. In \cite{2003_Takemura}, experimental observations were performed with fully contaminated spherical bubbles rising near a wall in a liquid at rest under conditions $\Rey\lesssim100$; such bubbles behave essentially as rigid torque-free spheres. Theoretical considerations about the nature of the particle-wall interaction were summarized through the semi-empirical expression for the transverse force coefficient
\begin{equation}
C_{\text Lu}^\text W [\Rey=\mathcal{O}(1-100)] \approx a^2(\Rey) (L_\text{R}/3)^{g(\Rey)}C_{\text Lu}^\text{W-out}(\Rey\lesssim1)\,,
\label{eq:CLuW_moderRe}
\end{equation}
with
\refstepcounter{equation}
$$
a(\Rey) = 1+0.6\Rey^{0.5}-0.55\Rey^{0.08} \quad\mbox{and}\quad
g(\Rey)=-2.0\tanh (0.01\Rey)\,.
\eqno{(\theequation{\text{a},\text{b}})}\label{eq:a0}
$$


\section{Flow field and fundamental mechanisms}\label{sec:4}


\subsection{Non-rotating sphere}\label{sec:4.1}	

\begin{figure}
	\centerline{\includegraphics[scale=0.85]{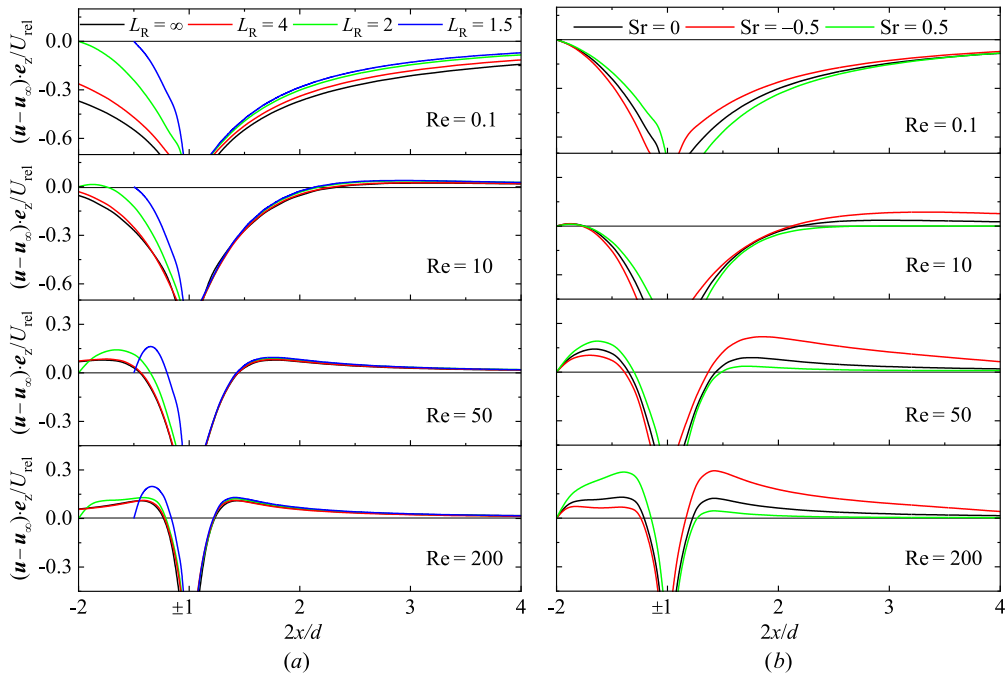}}
	\caption{
		Distribution of the streamwise velocity disturbance $ (\boldsymbol{u}-\boldsymbol{u}_\infty)\cdot  \boldsymbol{e}_z/U_\text{rel} $ along the $x$-axis (to magnify the flow in the gap, the $]-1,+1[$ part of the axis has been cut, so that the sphere is shrunk to a point). (a) Stagnant fluid ($\Sry = 0$) for different separation distances; (b) linear shear flow for $L_\text R=2$. The wall stands at $2x/d=-1.5$, $-2$, and $-4$ for $L_\text{R}=1.5$, $2$, and $4$, respectively.}
	\label{fig:disturbVel}
\end{figure}

Figure \ref{fig:disturbVel}(a) shows how the distribution of the streamwise velocity disturbance along the line ($y = 0$, $z = 0$) perpendicular to the wall, i.e. the $x$-axis, varies with flow conditions in the case of a sphere translating in a stagnant fluid. The sphere leading the fluid,  $U_\text{rel}$ is negative, so that negative (positive) normalized velocities correspond to an upward (downward) fluid motion. At high Reynolds number ($\Rey=200$), the no-slip condition induces a thin boundary layer around the sphere, within which the disturbance is always negative. Outside this boundary layer, the fluid is accelerated by the sphere motion, making the disturbance become positive on both sides. Owing to the finite-gap offered to the fluid, this acceleration is more pronounced on the wall-facing side and the maximum velocity increases as the wall-sphere separation decreases. In this high-$\Rey$ configuration, the velocity disturbance outside the boundary layer remains positive throughout the gap. 
Wall-proximity effects sharply decrease as the gap widens and are almost negligible for $L_\text R\gtrsim4$. 
 The boundary layer thickens as $\Rey$ decreases and viscous effects increasingly control the flow in the gap. For instance, the velocity disturbance keeps a negative sign throughout the gap for $L_\text{R}=1.5$ when $\Rey=10$, and only passes through a tiny positive maximum for $L_\text{R}=2$ before returning to zero at the wall. In such cases, the fluid in the gap is essentially entrained by the sphere translation. 
 For each $L_\text{R}$, the velocity disturbance at a given distance from the sphere surface is seen to reach larger negative values on the wall-facing side compared to the `free' side, illustrating the enhancement of viscous effects in the gap due to the nearby wall.

The influence of the shear on the disturbance flow is illustrated in Fig. \ref{fig:disturbVel}(b), based on the results obtained with two opposite relative shear rates, $\Sry=\pm 0.5$, for a separation distance $L_\text R=2$. In the moving frame, the presence of a positive (negative) shear, corresponding to the configuration where the sphere lags (leads) the fluid, accelerates (decelerates) the flow on the wall-facing side, while it decelerates (accelerates) it on the opposite side. Consequently, compared to the un-sheared situation, the fluid acceleration is enhanced (reduced) on the wall-facing side by a positive (negative) shear when $\Rey$ is large (\Rey=100 or 200 in Fig. \ref{fig:disturbVel}(b)), while the opposite takes place on the `free' side. 
At moderate Reynolds number ($\Rey=10$), shear-induced acceleration/deceleration effects remain significant within the boundary layer and extend beyond it ($x/d>1$) on the `free' side. The shear-induced asymmetry is still present throughout the flow at low Reynolds number ($\Rey=0.1$), the disturbance velocity remaining negative everywhere (i.e. directed upstream of the local carrying flow) in this case. 
\begin{figure}
	\centerline{\includegraphics[scale=0.85]{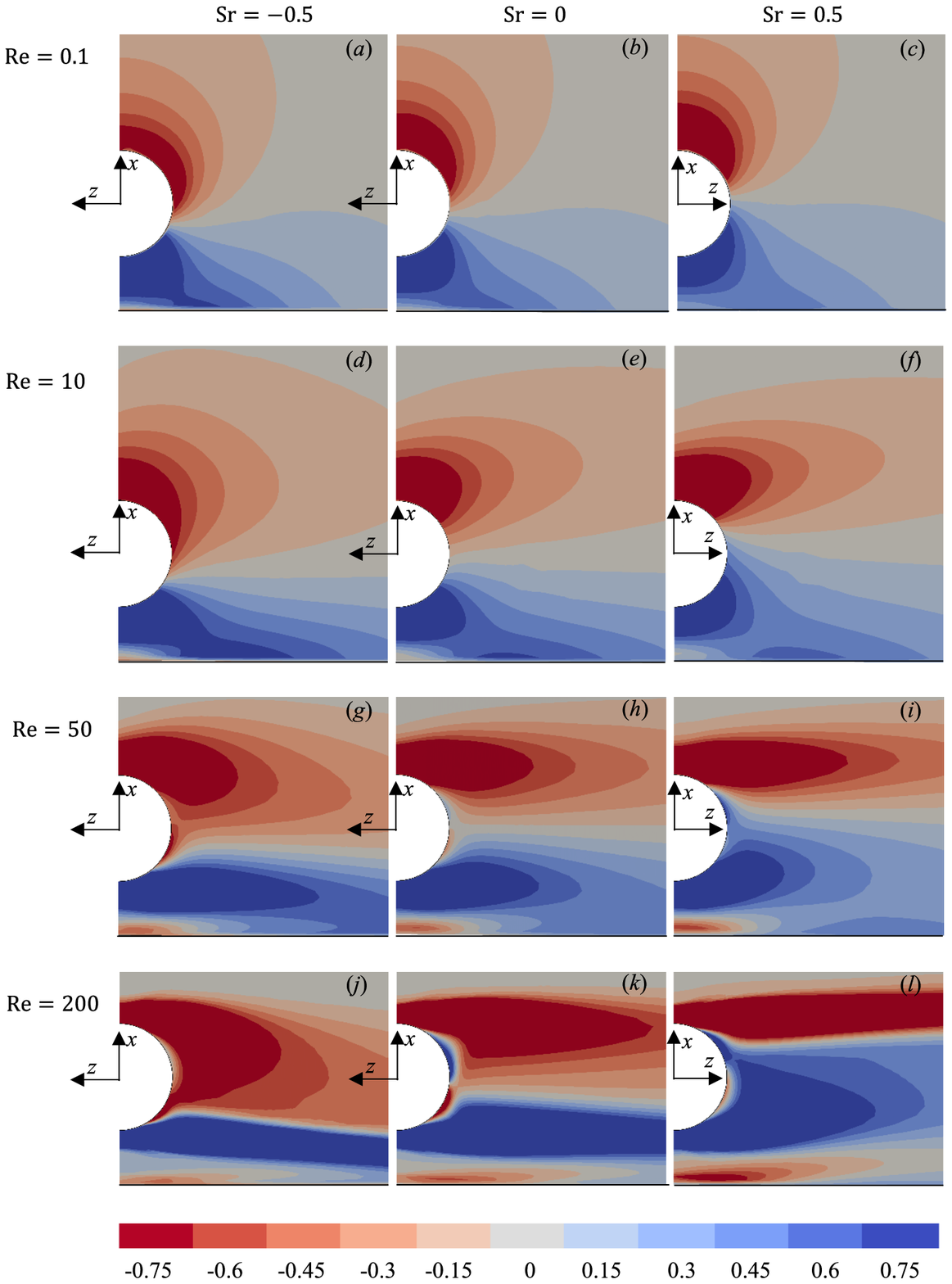}}
	\caption{
		Iso-contours of the normalized spanwise vorticity disturbance $d/(2U_\text{rel})(\boldsymbol{\omega}-\boldsymbol{\omega}_\infty) \cdot  \boldsymbol{e}_y$ in the symmetry plane $y = 0$ for $L_\text R=2$. Left column: $\Sry=-0.5$ ($U_\text{rel}<0$); central column: $\Sry=0$ ($U_\text{rel}<0$); right column: $\Sry=0.5$ ($U_\text{rel}>0$). The wall stands at the bottom of each panel. The relative flow with respect to the sphere is from left to right, i.e. in the $z$-direction for $\Sry=0.5$ and in the $-z$-direction for $\Sry=0$ and $-0.5$.}
	\label{fig:spanwiseVor}
\end{figure}

The distribution of the spanwise component of the vorticity disturbance in the symmetry plane $y = 0$ is displayed in Fig. \ref{fig:spanwiseVor} for the specific separation $L_\text{R}=2$. Vorticity is generated both at the sphere surface and at the wall, owing to  the no-slip condition on both surfaces. We refer to the corresponding two contributions in the vorticity field as the `surface' vorticity and `wall' vorticity, respectively. 
When the fluid is at rest at infinity, the surface vorticity is advected asymmetrically, preferentially towards the wall at high Reynolds number (Fig. \ref{fig:spanwiseVor}(k)). A thin layer of wall vorticity, the strength of which increases with $\Rey$, takes place in the lower part of the gap. 
In the same panel, it may be noticed that the stagnation point at the back of the sphere stands slightly below the plane $x=0$, i.e. it is shifted towards the wall compared to unbounded flow configuration, in agreement with previous observations \cite{2005_Zeng}. When the Reynolds number decreases, the thickness of the two boundary layers increases (e.g. $\Rey=50$ in Fig. \ref{fig:spanwiseVor}(h)), reinforcing their interaction. 
At lower Reynolds number ($\Rey=10$ and 0.1, Figs. \ref{fig:spanwiseVor}(e) and \ref{fig:spanwiseVor}(b)), diffusion in the crosswise ($x$) direction is sufficiently efficient to allow the surface vorticity to control the wall region, except in the narrowest part of the gap ($|z|/d\ll1$). In this regime, the vorticity distribution is essentially similar to that observed in \cite{2020_Shi_a} with a spherical bubble, up to a factor $3/2$ resulting from the difference in the magnitude of the Stokeslet (hence the drag force) associated with the two types of bodies. Thus, the mechanisms responsible for the drag enhancement and the transverse force are similar to those discussed in \cite{2020_Shi_a}. 
In particular, the gradual slowing down of the fluid displaced by the sphere along the wall as the downstream distance increases induces a small transverse flow correction directed away from the wall, which is responsible for the repulsive transverse force acting on the sphere.\\
\indent In the presence of an ambient shear, a shear-flow type correction has to take place within the boundary layer for the no-slip condition to be satisfied at the sphere surface, yielding a negative correction in the spanwise vorticity therein when the sphere lags the fluid. Hence this correction enhances the primary negative vorticity on the `free' side ($x>0$), while it lowers the primary positive vorticity in the part of the boundary layer facing the wall ($x<0$), as Fig. \ref{fig:spanwiseVor}(i) confirms. The process reverses when the sphere leads the fluid, in agreement with Fig. \ref{fig:spanwiseVor}(g). The wall vorticity in the gap is also modified by the shear: it increases (decreases) for $\Sry>0$ ($\Sry<0$), owing to the acceleration (deceleration) of the fluid on the wall-facing side caused by the positive (negative) shear, as Figs. \ref{fig:spanwiseVor}(l) and (j) confirm. 
\begin{figure}
	\centerline{\includegraphics[scale=1]{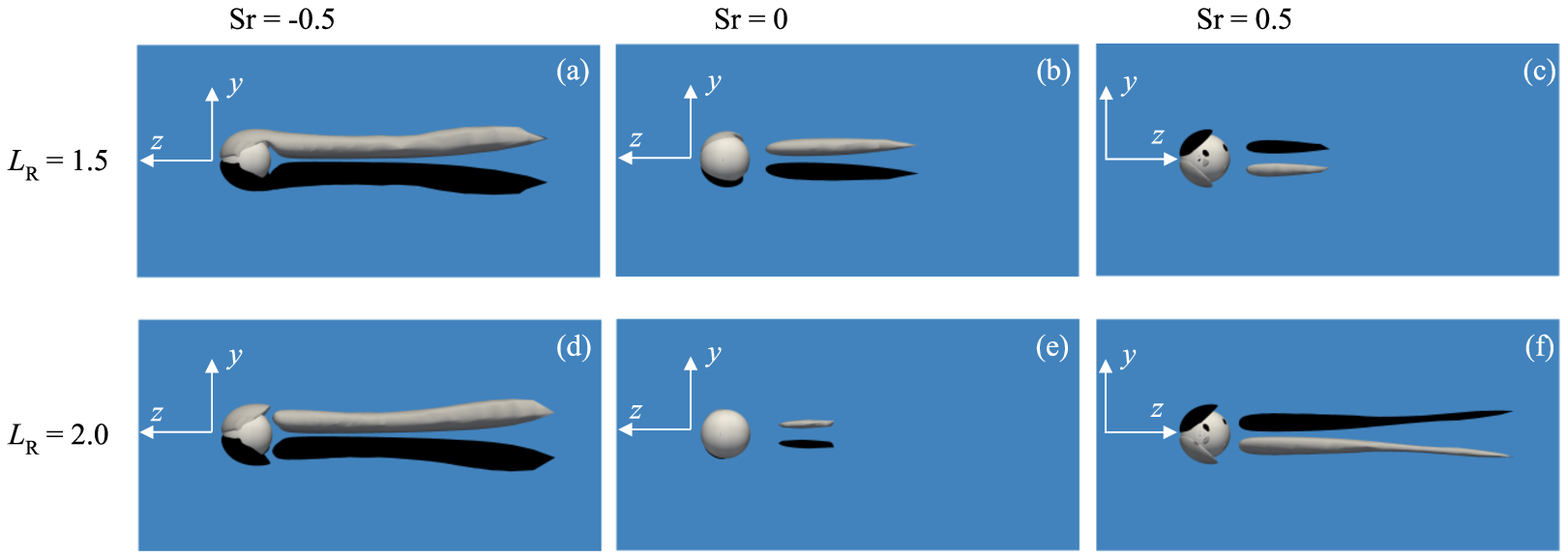}}
	\caption{
		Iso-surfaces $d/(2|U_\text{rel}|)\boldsymbol{\omega} \cdot  \boldsymbol{e}_z=\pm0.25$ of the streamwise vorticity in the wake of a sphere moving parallel to a wall at $\Rey = 200$ (the black thread corresponds to positive values). Left column: $\Sry=-0.5$ ($U_\text{rel}<0$); central column: $\Sry=0$ ($U_\text{rel}<0$); right column: $\Sry=0.5$ ($U_\text{rel}>0$). The flow with respect to the sphere is from left to right, i.e. in the $z$-direction for $\Sry=0.5$ and in the $-z$-direction for $\Sry=0$ and $-0.5$. Since $(x,y,z)$ is right-handed, the wall stands at the back of the sphere for $\Sry=0.5$ and at its back for $\Sry=0$ and $-0.5$.}
	\label{fig:streamwise_Vor_wb}
\end{figure}

 The near-wall situation makes the flow past the sphere intrinsically three-dimensional, even when the fluid is at rest at infinity. Consequently, the streamwise component of the vorticity, $\omega_z=\boldsymbol{\omega} \cdot  \boldsymbol{e}_z$, is nonzero in the wake, unlike in the axisymmetric configuration prevailing in the unbounded case at low and moderate $\Rey$. The $\omega_z$-distribution in the un-sheared case is shown in Figs. \ref{fig:streamwise_Vor_wb}(b) and \ref{fig:streamwise_Vor_wb}(e) at $\Rey = 200$ for the two separations $L_\text{R}=1.5$ and $L_\text{R}=2.0$, respectively. 
  The streamwise vorticity is concentrated within two elongated vortices standing on both sides of the symmetry plane $y=0$. The fluid located in between the two vortex threads is entrained downwards, bending the iso-contours of the spanwise vorticity towards the wall, as seen in Fig. \ref{fig:spanwiseVor}(k). Three-dimensional effects sharply decrease as $L_\text{R}$ increases, and so does the strength of $\omega_z$ as shown by Fig. \ref{fig:streamwise_Vor_wb}(e). In an unbounded flow, the axial symmetry in the wake of a sphere is known to break down at a critical Reynolds number $\Rey^{SS}\approx212.6$ through a stationary bifurcation \citep{1993_Natarajan,2012_Fabre}, leading to a stationary flow with a double-threaded wake structure qualitatively similar to that depicted in Fig. \ref{fig:streamwise_Vor_wb}(b), and a symmetry plane whose orientation is selected by some initial disturbance.
\begin{figure}
	\centerline{\includegraphics[scale=1]{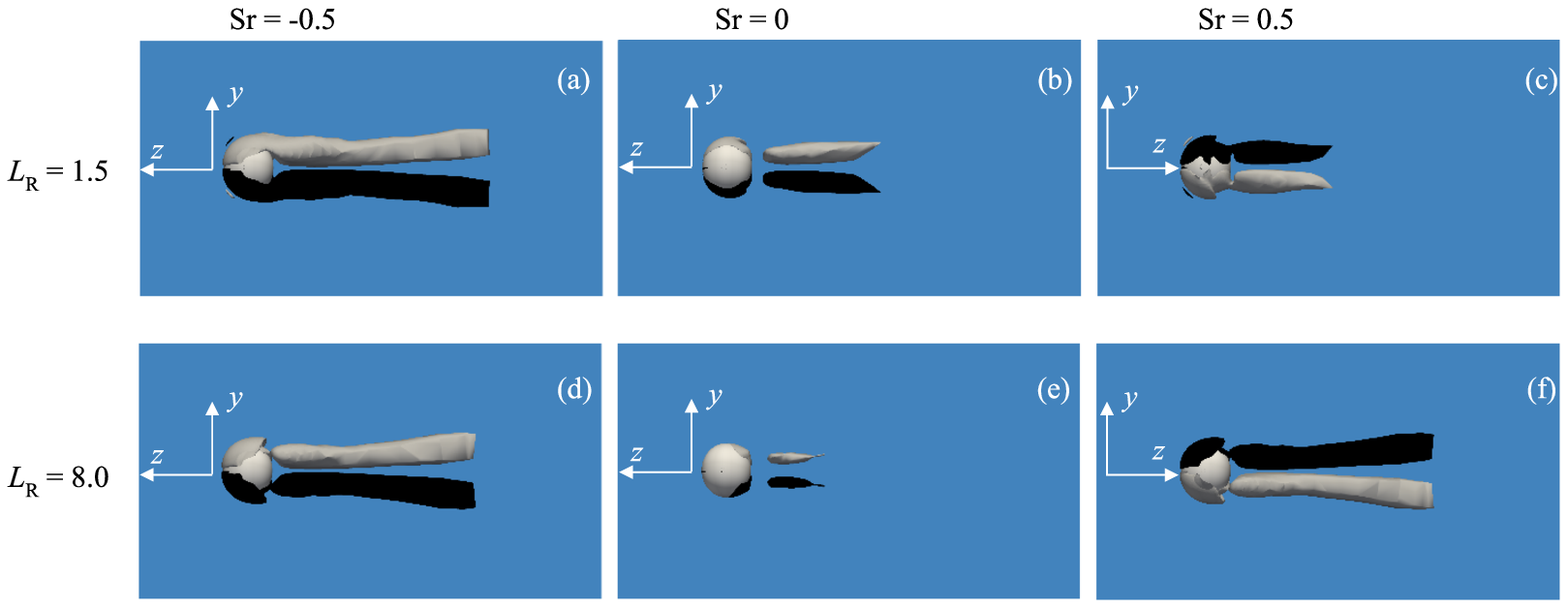}}
	\caption{
		Same as Fig. \ref{fig:streamwise_Vor_wb} for $\Rey = 250$.}
	\label{fig:streamwise_Vor_wb2}
\end{figure}
In the presence of a nearby wall, the flow structure observed for $\Rey\gtrsim \Rey^{SS}$ results from the combination of the above two mechanisms, the presence of the wall dictating the orientation of the symmetry plane \citep{2005_Zeng}. The corresponding wake structure is illustrated in Fig. \ref{fig:streamwise_Vor_wb2}(e) for $\Rey = 250$ and two separations, $L_\text{R}=1.5$ and $8$. Now, the strength of the streamwise vortices is significant even for $L_\text{R}=8$, owing to the intrinsic instability of the axisymmetric wake. At such large separations and Reynolds number, the sign of the streamwise vorticity in each vortex thread is dictated by the slight acceleration of the fluid in the gap: based on Bernoulli's theorem, this acceleration is seen to imply a pressure minimum there, forcing the fluid located within the symmetry plane $y=0$ (i.e. in between the two streamwise vortices) to be deviated toward the wall. Continuity then implies that the fluid must go away from the wall on the outer side of the streamwise vortices, yielding a transverse force toward $x>0$. Consequently, the wall-interaction and the intrinsic wake instability mechanisms cooperate when the separation distance decreases, enhancing the strength of the streamwise vortices, as the comparison between Fig. \ref{fig:streamwise_Vor_wb2}(b) and Figs. \ref{fig:streamwise_Vor_wb2}(e) and \ref{fig:streamwise_Vor_wb}(b) confirms. 

\begin{figure}
	\centerline{\includegraphics[scale=0.85]{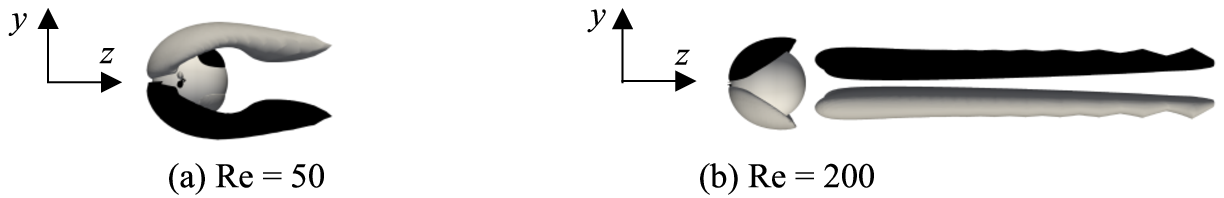}}
	\caption{
		Iso-surfaces of the streamwise vorticity in the wake of a sphere moving in an unbounded linear shear flow with $\Sry = 0.5$ (the black thread corresponds to the positive value). (a) $\Rey=50$, $d/(2|U_\text{rel}|)\boldsymbol{\omega} \cdot  \boldsymbol{e}_z=\pm0.1$; (b) $\Rey = 200$, $d/(2|U_\text{rel}|)\boldsymbol{\omega} \cdot  \boldsymbol{e}_z=\pm0.25$. The relative upstream flow is from left to right and the free vorticity $-\gamma$ lies along the $y$-direction. 
		}
	\label{fig:streamwise_Vor_ub}
\end{figure}

In the presence of a mean shear, the `free' vorticity $\boldsymbol{\omega}_\infty=\nabla\times\boldsymbol{u}_\infty=-\gamma \boldsymbol{e}_ {y}$ comes into play. In an unbounded flow domain, it yields the classical shear-induced lift force associated with the lift coefficient (\ref{eq:U_low_Re}) in the low-$\Rey$ regime. In the case of a rigid non-rotating sphere, a remarkable feature is that this force changes sign for $\Rey\gtrsim50$ \citep{1999_Kurose,2002_Bagchi}, mostly because of the nearly-uniform shear stress distribution within the recirculation attached to the rear part of the sphere. This change of sign, which follows that of the streamwise vorticity within each vortex thread, is confirmed in Fig. \ref{fig:streamwise_Vor_ub}.
 In the presence of a nearby wall, a consequence of this change of sign is that, provided $\Rey\gtrsim50$, the shear-induced and slip-induced mechanisms cooperate when $\Sry$ is negative and act in an antagonistic manner when $\Sry$ is positive, while the reverse happens for $\Rey\lesssim50$. The influence of the sign of $\Sry$ in the former case is confirmed in Fig. \ref{fig:streamwise_Vor_wb} (\Rey=200), since the trailing vortices observed when $\Sry<0$ (Figs. \ref{fig:streamwise_Vor_wb}(a) and \ref{fig:streamwise_Vor_wb}(d)) are thicker than in the unbounded case (Fig. \ref{fig:streamwise_Vor_ub}(b)), while they are thinner when $\Sry>0$ (Figs. \ref{fig:streamwise_Vor_wb}(c) and \ref{fig:streamwise_Vor_wb}(f)). 
  The presence of the double-threaded wake and the variation of its strength with the sign of $\Sry$ have a direct influence on the advection of the surface vorticity downstream of the sphere. Indeed, according to the direction of the streamwise vorticity in each vortex thread, this wake structure entrains the fluid standing close to the mid-plane $y=0$ towards (away from) the wall when $\Sry<0$ ($\Sry>0$). Since the streamwise vortices are stronger in the former case, so is the resulting bending of the wake towards the wall (Fig. \ref{fig:spanwiseVor}(j)), as compared to its bending toward the fluid interior when $\Sry$ is positive (Fig. \ref{fig:spanwiseVor}(l)).\\ 
\begin{figure}
	\centerline{\includegraphics[scale=1]{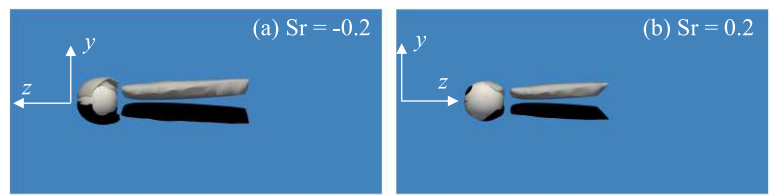}}
	\caption{Same as Fig. \ref{fig:streamwise_Vor_wb2}(a) and (c) for $\Sry=\pm0.2$.}
	\label{fig:streamwise_Vor_wb3}
\end{figure}
\indent For supercritical Reynolds numbers, i.e. $\Rey>\Rey^{SS}$, the above picture still holds when $\Sry$ is negative, since all mechanisms involved in the generation of the streamwise vorticity cooperate. The only difference is that the magnitude of $\omega_z$ is increased compared to subcritical conditions, since the wake instability contributes to reinforce this vorticity component (compare the diameters of the $\omega_z$-iso-surfaces corresponding to $\Rey = 200$ and $250$ in Figs. \ref{fig:streamwise_Vor_wb}(a) and \ref{fig:streamwise_Vor_wb2}(a)).
In contrast, when $\Sry$ is positive, the mechanism associated with the free vorticity and those related to the wall proximity and wake instability act in an antagonistic manner. Therefore, the resulting sign of the streamwise vorticity in each vortex thread depends on the magnitude of $\Sry$. For large enough relative shear rates, this sign follows that found in the unbounded configuration. As the comparison between Figs. \ref{fig:streamwise_Vor_wb2}(c) and \ref{fig:streamwise_Vor_ub}(b) shows, this is the case with $\Sry=0.5$ at $\Rey=250$. 
In contrast, mechanisms related to the wall proximity and wake instability dominate when the ambient shear is weak enough. This situation is illustrated in Fig. \ref{fig:streamwise_Vor_wb3}(b) ($\Sry = 0.2$), where the sign of $\omega_z$ in each vortex thread is seen to be opposite to that found in Fig. \ref{fig:streamwise_Vor_wb2}(c) with $\Sry=0.5$ at the same Reynolds number and separation from the wall.\\
\indent As the Reynolds number decreases, shear-induced advective effects in the wake weaken and vorticity diffusion across the wall-particle gap becomes increasingly important. 
For instance, bending of the surface vorticity toward or away from the wall is no longer observed in Figs. \ref{fig:spanwiseVor}(g-i) at $\Rey=50$. At $\Rey = 10$, the boundary layer is thick enough for  the positive vorticity disturbance generated on the wall-facing part of the sphere surface to interact directly with the negative wall vorticity disturbance, similar to the un-sheared case. As $\Rey$ further decreases, viscous diffusion becomes so strong that the surface vorticity virtually controls the entire wall region. Nevertheless, influence of the ambient  shear is still present, favoring (reducing) the diffusion of the surface vorticity toward the wall when $\Sry$ is negative (positive), as the iso-contours in Figs. \ref{fig:spanwiseVor}(a-c) reveal. The surface vorticity being enhanced (reduced) on the wall-facing side for negative (positive) $\Sry$, the drag acting on the sphere is increased (reduced), which is reflected in the last term in the right-hand side of \eqref{eq:CDWin_tf}. The wall- and shear-induced mechanisms both yield a transverse force directed toward $x > 0$ if $\Sry$ is positive in the low-but-finite $\Rey$ regime. Hence they act together to produce an enhanced repulsive force in this configuration, as reflected in (\ref{eq:CLWin}), whereas their antagonistic action yields a reduced transverse force when $\Sry$ is negative. 
\subsection{Torque-free sphere}\label{sec:4.2}	
\begin{figure}
	\centerline{\includegraphics[scale=0.85]{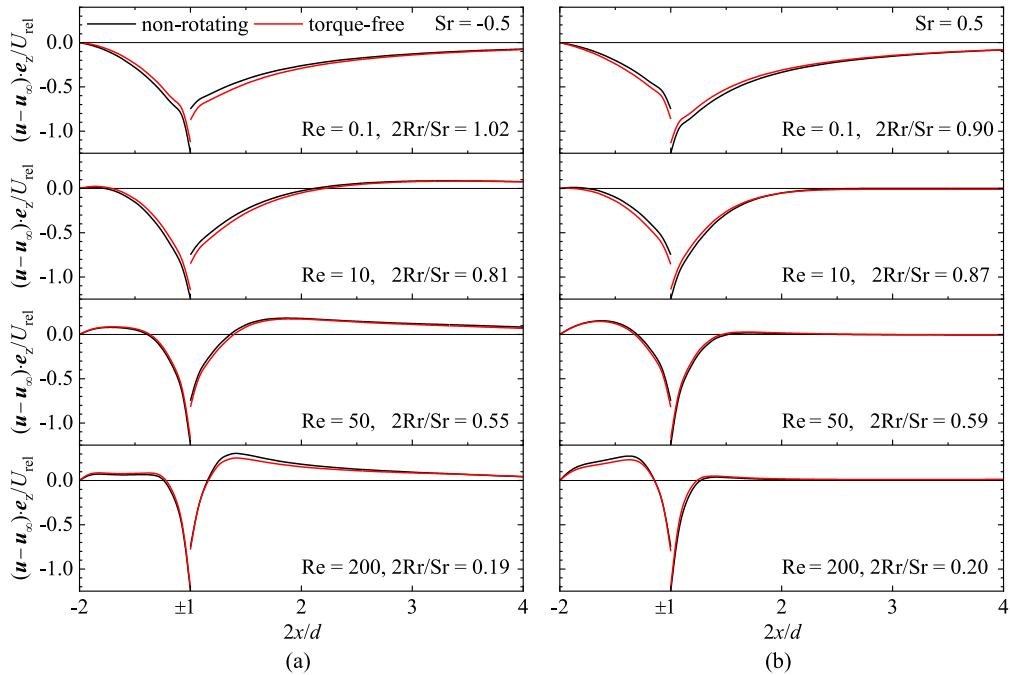}}
	\caption{
		Influence of the sphere rotation on the distribution of the streamwise velocity disturbance $(\boldsymbol{u}-\boldsymbol{u}_\infty)\cdot  \boldsymbol{e}_z/U_\text{rel}$ along the $x$-axis for $L_\text R=2$. (a) $\Sry = -0.5$; (b) $\Sry = 0.5$. The wall stands at position $2x/d=-2$ and the interval $]-1\leq2x/d\leq+1[$ has been cut. The magnitude of the normalized sphere rotation rate $2\Rry/\Sry=2\Omega/\gamma$ in the torque-free case is indicated in each panel.}
	\label{fig:disturbVel_tf}
\end{figure}
\indent Figure \ref{fig:disturbVel_tf} compares the profiles of the streamwise velocity disturbance along the $x$-axis in the case of a torque-free sphere with that of a non-rotating sphere, both with $\Sry=\pm 0.5$ and $L_\text R=2$. Values of the normalized rotation velocity $2\Rry/\Sry=2\Omega/\gamma$ indicated in each panel reveal a significant decrease of the rotation rate as $\Rey$ increases, $\Rry$ typically reducing by a factor of five from $\Rey=0.1$ to $\Rey=200$. A similar tendency has been reported in the unbounded case \citep{2002_Bagchi,2013_Homann}. Rotation being clockwise for $\Sry>0$, the streamwise velocity is found to decrease on the wall-facing side and increase on the opposite side; the reverse happens when $\Sry$ is negative. However the corresponding changes are minimal and vanish beyond a distance to the sphere surface of the order of its radius. Analyzing the spatial distribution of the spanwise vorticity disturbance (not shown) leads to the same conclusion.\\
\begin{figure}
	\centerline{\includegraphics[scale=0.85]{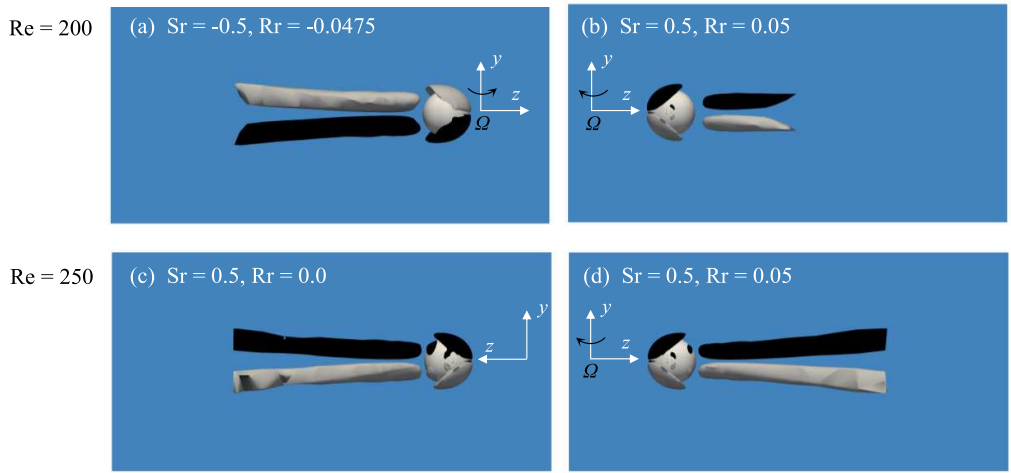}}
	\caption{
		Iso-surfaces $d/(2|U_\text{rel}|)\boldsymbol{\omega} \cdot  \boldsymbol{e}_z=\pm0.25$ of the streamwise vorticity around a torque-free sphere for $L_\text R=2$ and $|\Sry| = 0.5$. 
}
	\label{streamwise_Vor_wb_tf}
\end{figure}
\indent Things differ at high Reynolds number for the streamwise vorticity component. Figure \ref{streamwise_Vor_wb_tf} shows the structure of the $\omega_z$-field in the wake of a torque-free sphere for $\Sry=\pm0.5$ and a separation distance $L_\text{R}=2$. The aforementioned rotation-induced changes in the fluid velocity at the particle surface lower the actual shear `felt' by the sphere whatever the sign of $\Sry$. Therefore the source term responsible for the generation of the streamwise vorticity is lowered by the rotation, resulting in a weaker pair of vortex threads compared to the non-rotating configuration. Comparing Fig. \ref{streamwise_Vor_wb_tf} with its counterpart in the case of a non-rotating sphere (Figs. \ref{fig:streamwise_Vor_wb}(d) and (f)) confirms this conclusion. 
In contrast, under supercritical conditions, the generation of $\omega_z$  for similar levels of $|\Sry|$ is essentially governed by the wake instability, not by the shear around the particle. Consequently, little change is expected between the non-rotating and torque-free configurations, which Figs. \ref{streamwise_Vor_wb_tf}(c)-(d) confirm. 

\section{Hydrodynamic forces}\label{sec:5}	
We now discuss the variations of the computed drag and lift forces acting on the sphere with the various control parameters. Most results were obtained by considering the parameter range $0.1\le \Rey\le250$ and $|\Sry|\le0.5$, within which the flow field in the particle frame is steady for all considered $L_\text{R}$. Numerical data are systematically used to derive empirical or semiempirical force models. Most of these models are valid only within a specific $\Rey$-range but we frequently combine them to obtain empirical fits valid throughout the considered range of Reynolds number.

\subsection{Fluid at rest at infinity}\label{sec:5.1}
\subsubsection{Drag}\label{sec:5.1.1}

\begin{figure}
	\centerline{\includegraphics[scale=0.9]{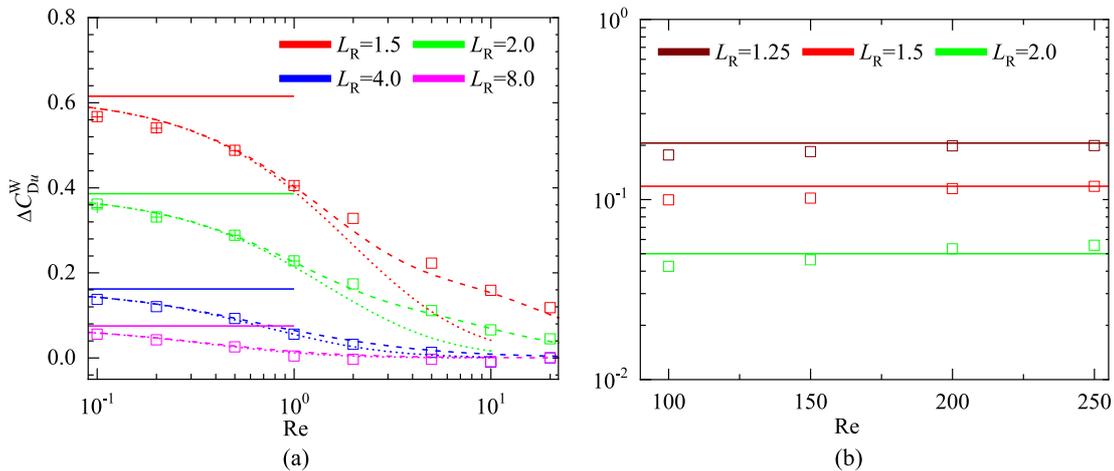}}
	\caption{Relative wall-induced drag increase $\Delta C_{\text Du}^\text W$ for a sphere moving parallel to a wall in a fluid at rest. (a) Low-to-moderate-$\Rey$ regime. (b) High-$\Rey$ regime.
 $\square$ and $+$: numerical data corresponding to a non-rotating and a torque-free sphere, respectively; solid lines in (a): zero-$\Rey$ asymptotic prediction \eqref{eq:CDWin_tf}; dotted lines: low-but-finite-$\Rey$ prediction (\ref{drag_fin}); dashed lines: low-to-moderate-$\Rey$ semiempirical prediction \eqref{eq:Wu-drag-low-to-mo-Re}; solid lines in (b): high-$\Rey$ expression \eqref{eq:Wu-drag-high-Re}.}.
	\label{fig:drag_sr0_low-and-high-Re}
\end{figure}

It is known since Faxén's pioneering work \citep{1965_Happel} that the presence of a wall increases the drag in the low-$\Rey$ limit. This increase, say, $\Delta C_{\text Du}^\text W=(C_{\text Du}^\text W-C_{\text D0}^\text U)/C_{\text D0}^\text U$, with $C_{\text D0}^\text U$ the drag coefficient on a sphere translating in an unbounded uniform flow, is displayed in Fig. \ref{fig:drag_sr0_low-and-high-Re}(a) for $0.1\leq \Rey \leq 20$ and various separation distances. Most results were obtained by considering a non-rotating sphere but data corresponding to the torque-free condition are also included for $\Rey\leq1$. No discernible difference is found between the two configurations, confirming the vanishingly small effect of the rotation induced by the torque-free condition on the drag in the range of separations considered here, in line with a previous remark on the asymptotic prediction (\ref{eq:CDWin_tf}). Numerical results closely approach this prediction (solid lines in Fig. \ref{fig:drag_sr0_low-and-high-Re}(a)) at $\Rey=0.1$. Inertial effects become increasingly important as the Oseen-length-based separation $L_u$ increases, making the drag increase depart from (\ref{eq:CDWin_tf}). The decrease in $\Delta C_{\text Du}^\text W$ as $\Rey$ increases is well captured by the low-but-finite-$\Rey$ expression \eqref{drag_fin} up to $\Rey = 1$. Following \cite{2003_Takemura}, this expression may be extended semi-empirically to moderate Reynolds numbers by noting that the drag increase in this regime is proportional to the square of the maximum surface vorticity. Variations of this quantity with $\Rey$ based on the results of  \cite{1995_Magnaudet} are expressed by the fitting function $a(\Rey)$ in (\ref{eq:a0}a). We performed specific runs in an unbounded uniform flow to check this expression and found that, for $\Rey\lesssim10$, these variations are more accurately approached by the fit $a(\Rey) \approx (1-0.12\Rey^{1/2}+0.37\Rey)^{1/2}$ which recovers the leading-order $0.6\Rey^{1/2}$-term of (\ref{eq:a0}a) at high Reynolds number. 
However, compared with the unbounded situation, 
the presence of a nearby wall tends to decrease the surface vorticity on the wall-facing side, as Fig. \ref{fig:spanwiseVor}(e) indicates. 
For this reason, we found that a more accurate estimate of the variations of the maximum surface vorticity in the near-wall configuration at moderate $\Rey$ is provided by   
\begin{equation}
a^\text W(\Rey,L_\text{R}) \approx \left\{1+\text{tanh}(0.05\Rey L_\text{R}^2)(0.37\Rey-0.12\Rey^{1/2})\right\}^{1/2}\,.
\label{eq:aW}
\end{equation}
Making use of \eqref{eq:aW}, which tends toward the above expression for $a(\Rey)$ at large distances from the wall, the low-but-finite-$\Rey$ wall-induced drag correction (\ref{drag_fin}) may be extended toward moderate Reynolds numbers in the form
\begin{equation}
\Delta C_{\text Du}^\text W[\Rey\lesssim10)]\approx f_\text D^{\prime}(L_u) [a^\text W(\Rey,L_\text{R})]^2 \Delta C_{\text Du}^\text{W-in}(L_\text{R})\,,
\label{eq:Wu-drag-low-to-mo-Re}
\end{equation}
where $\Delta C_{\text Du}^\text{W-in}$ corresponds to the low-$\Rey$ asymptotic prediction \eqref{eq:CDWin_tf} for $\Sry=0$ and $f_\text D^{\prime}$ is given by \eqref{eq:fD_prime}. 
As the dashed lines in Fig. \ref{fig:drag_sr0_low-and-high-Re}(a) show, \eqref{eq:Wu-drag-low-to-mo-Re} accurately captures the variations of $\Delta C_{\text Du}^\text W$ revealed by the simulations whatever $L_\text{R}$ up to $\Rey = 20$.

According to \eqref{eq:Wu-drag-low-to-mo-Re}, the wall-induced drag increase is vanishingly small beyond $\Rey \approx 100$. However, as Fig. \ref{fig:drag_sr0_low-and-high-Re}(b) reveals, numerical data in this regime indicate that this increase is still significant when the particle is close enough to the wall. 
Within the considered Reynolds number range ($100\leq \Rey\leq250$), this increase is found to depend only weakly on $\Rey$. In contrast, it varies dramatically with the inverse of the separation distance, increasing from $5\%$ for $L_\text{R}=2$ to $20\%$ for $L_\text{R}=1.25$. 
Fitting the results corresponding to $\Rey = 250$ yields
\begin{equation}
\Delta C_{\text Du}^\text W [\Rey=\mathcal{O}(100)] \approx 0.4L_\text{R}^{-3}\,.
\label{eq:Wu-drag-high-Re}
\end{equation}
Figure \ref{fig:drag_sr0_low-and-high-Re}(b) shows that \eqref{eq:Wu-drag-high-Re} captures the observed drag increase well for $\Rey\gtrsim100$. 
The $-3$ exponent in \eqref{eq:Wu-drag-high-Re} is readily understood by noting that there is little direct interaction between the near-sphere and near-wall vortical regions in this regime (see Fig. \ref{fig:spanwiseVor}(k)). Therefore, the sphere-wall interaction has an almost inviscid nature, meaning that the sphere perceives the wall essentially as a free-slip plane and the latter perceives the sphere as an irrotational dipole (associated with its finite size). The image dipole required to satisfy the non-penetration condition on a nearby plane is known to induce an $\mathcal{O}(L_\text{R}^{-3})$-increase in the relative velocity of the fluid at the sphere center, which in turn increases the viscous dissipation resulting from the sphere motion by a similar amount \citep{2003_Legendre}. Equating the dissipation rate with the rate of work of the drag force then implies that $\Delta C_{\text Du}^\text W$ is proportional to $L_\text{R}^{-3}$.
\begin{figure}
	\centerline{\includegraphics[scale=0.9]{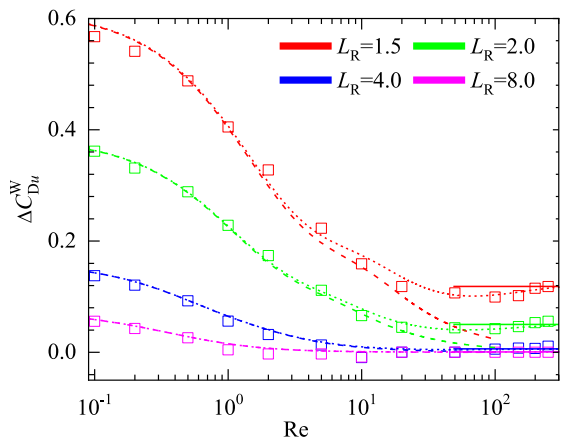}}
	\caption{Wall-induced drag correction $\Delta C_{\text Du}^\text W$ for a non-rotating sphere translating parallel to a wall in a fluid at rest in the range $0.1\leq\Rey\leq250$. Symbols: numerical data; solid lines: high-$\Rey$ correction (\ref{eq:Wu-drag-high-Re}); dashed lines: low-to-moderate-$\Rey$ correction \eqref{eq:Wu-drag-low-to-mo-Re}; dotted lines: composite fit (\ref{eq:Wu-drag}).}
	\label{fig:drag_sr0_low-to-high-Re}
\end{figure}

Figure \ref{fig:drag_sr0_low-to-high-Re} compares the predictions provided by expressions (\ref{eq:Wu-drag-low-to-mo-Re}) and (\ref{eq:Wu-drag-high-Re}) with the numerical data obtained throughout the $\Rey$-range investigated. Obviously none of them is appropriate in the intermediate range $20\lesssim \Rey\lesssim100$. For practical purposes, an empirical fit resulting from the combination of the two models is desirable. Noting that the drag excess predicted by (\ref{eq:Wu-drag-low-to-mo-Re}) becomes vanishingly small when the Reynolds number exceeds a few tens, a linear combination of (\ref{eq:Wu-drag-low-to-mo-Re}) and (\ref{eq:Wu-drag-high-Re}) with a suitable pre-factor of the latter ensuring that its effect vanishes at low Reynolds number appears to be convenient. Calibrating this pre-factor in the intermediate $\Rey$-range, we obtained
\refstepcounter{equation}
$$
\Delta C_{\text Du}^\text W(\Rey)\approx\Delta C_{\text Du}^\text W[\Rey\lesssim10]+c_{Du\infty}(\Rey)\Delta C_{\text Du}^\text W[\Rey=\mathcal{O}(100)]\,, \quad\mbox{with}\quad  c_{Du\infty}(\Rey)=1-\exp^{-0.035\Rey^{0.75}}\,.
\eqno{(\theequation{\text{a},\text{b}})}
\label{eq:Wu-drag}
$$
As the dotted lines in Fig. \ref{fig:drag_sr0_low-to-high-Re} show, this composite expression correctly reproduces the observed wall-induced drag increase whatever the Reynolds number.

\subsubsection{Transverse force}\label{sec:5.1.2}	

\begin{figure}
	\centerline{\includegraphics[scale=0.9]{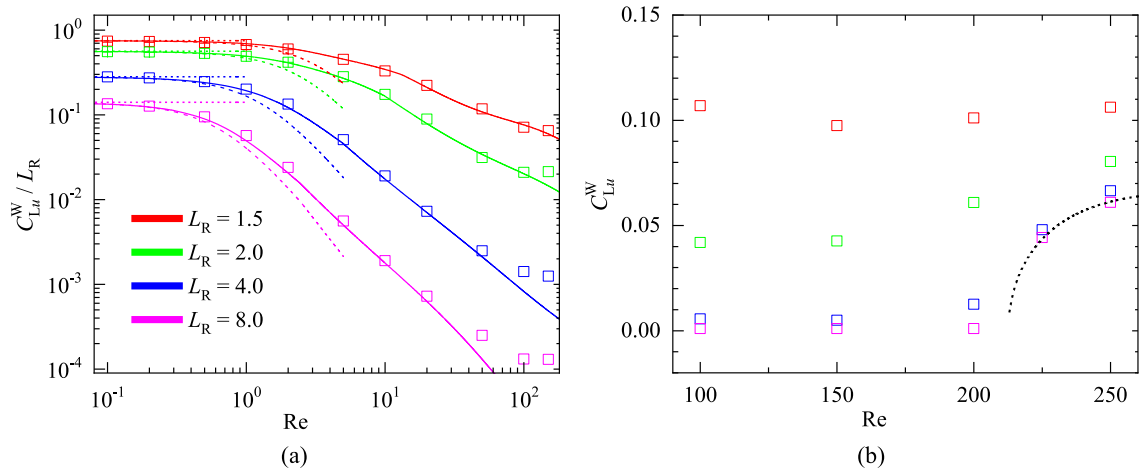}}
	\caption{Lift coefficient (divided by $L_\text{R}$ in (a) for a better readability) on a sphere translating parallel to a wall in a fluid at rest. (a) regime $0.1\leq\Rey\leq100$. (b) high-$\Rey$ regime $\Rey\geq100$. $\square$ and $+$: numerical data for a non-rotating and a torque-free sphere, respectively; dotted lines in (a): leading-order term of asymptotic expressions (\ref{eq:CLWin}) and (\ref{eq:CLWin_tf}); dashed lines: approximate low-but-finite expression \eqref{eq:fL_prime}; solid lines: low-to-moderate $\Rey$ prediction \eqref{eq:CLuW_moderRe} with $a(\Rey)$ substituted with $a^W(\Rey)$ as given by (\ref{eq:aW}); dotted line in (b): prediction \eqref{eq:CL_Bi} for the lift force in an unbounded fluid beyond the stationary bifurcation.}
	\label{fig:lift_sr0_low-and-high-Re}
\end{figure}

Figure \ref{fig:lift_sr0_low-and-high-Re}(a) shows the numerical data obtained for the wall-induced transverse force in the range $0.1\leq \Rey\leq150$ for a non-rotating sphere. Data corresponding to a torque-free sphere obtained at small wall distances and Reynolds numbers less than unity are also shown. Given the vanishingly small difference between the results corresponding to these two configurations at small $\Rey$ and the similar behavior observed in the moderate-to-high $\Rey$ regime in \cite{2005_Zeng}, it may be concluded that the sphere rotation associated with the torque-free condition has a negligible effect on the wall-induced transverse force. Although this effect is reflected in the difference among the pre-factors involved in \eqref{eq:CLWin} and \eqref{eq:CLWin_tf}, the overall difference between the two predictions amounts only to $0.3\%$ for $L_\text{R}=1.5$ and $0.4\%$ for $L_\text{R}=1.25$, confirming the above statement. Numerical results at $\Rey=0.1$ are in good agreement with these asymptotic predictions, beyond which the transverse force gradually decreases as inertial effects increase, making the wall move from the inner region of the disturbance to the outer region. This decrease is well captured by \eqref{eq:fL_prime} up to $\Rey\approx1$. Further increasing $\Rey$ reveals that the force predicted by this low-but-finite $\Rey$ approximation decreases too fast. A similar trend was noticed in \cite{2003_Takemura}, leading to the semiempirical extension \eqref{eq:CLuW_moderRe} of the previous prediction. This extension was obtained by noting that, similar to the wall-induced drag increase, the transverse force at low-to-moderate $\Rey$ is proportional to the square of the maximum vorticity at the sphere surface, and the dependence of this force with respect to $L_u$ varies from $L_u^{-2}$ for $\Rey\lesssim1$ to $L_u^{-4}$ for $\Rey\gg1$. In \cite{2005_Zeng} and \cite{2009_Zeng}, this extended prediction was found to be in good agreement with numerical results up to $L_u=100$ for $1.5\leq L_\text{R}\leq8$.
Predictions of \eqref{eq:CLuW_moderRe}, with the slight change from $a(\Rey)$ to $a^\text{W}(\Rey)$ as given in \eqref{eq:aW} are shown in Fig. \ref{fig:lift_sr0_low-and-high-Re}(a). They are seen to capture the variations of the transverse force well up to $\Rey\approx100$ for $L_\text{R}<4$. At larger $L_\text{R}$, they tend to underestimate the actual force for $\Rey\gtrsim20$. However, under such conditions, the residual values of the force are less than $1\%$ of the low-$\Rey$ value, making this underestimate of little significance. \\
\indent Figure \ref{fig:lift_sr0_low-and-high-Re}(b) shows how $C_{\text Lu}^\text{W}$ behaves for $\Rey\geq100$. For large enough wall-particle separations, typically $L_\text{R}\gtrsim4$, the transverse force is virtually zero up to $\Rey\approx200$. In this situation, the sphere is immersed in an almost uniform flow, so that its wake is essentially axisymmetric. The axial symmetry breaks down when the Reynolds number exceeds the critical value $\Rey=\Rey^{SS}$, giving rise to a nonzero transverse force at larger Reynolds numbers. The wall plays no role in the occurrence of this force, nor in its intensity. However it selects the orientation of the symmetry plane that characterizes the wake beyond the threshold Reynolds number, hence that of the transverse force, according to the mechanism discussed in Sec. \ref{sec:4.1}. The corresponding imperfect bifurcation being supercritical and of pitchfork type \cite{2016_Citro}, it gives rise to a force growing as the square root of $\Rey-\Rey^{SS}$ close to the threshold. The associated pre-factor ($\approx2.95$) was computed exactly through a weakly nonlinear approach in the case of a translating sphere subject to a slow rotation in a fluid at rest, this rotation being responsible for the imperfect nature of the bifurcation \citep{2016_Citro}. An empirical extension of this prediction to non-negligible $\Rey-\Rey^{SS}$ gaps was proposed in \citep{2019_Shi} in the form
\begin{equation}
C_\text{L}^{\Rey>\Rey^{SS}} \approx2.95 \left(\frac{\Rey^{SS}}{\Rey}\right)^{1.5}\left[(\Rey^{SS})^{-1}-\Rey^{-1}\right]^{1/2} \,.
\label{eq:CL_Bi}
\end{equation}
Figure \ref{fig:lift_sr0_low-and-high-Re}(b) shows that present results corresponding to $L_\text{R}=8$ follow closely this prediction up to $\Rey=250$, i.e. nearly $20\%$ beyond the threshold. As the wall-sphere separation decreases, the situation becomes less clear-cut because the flow `felt' by the sphere remains significantly anisotropic, even for $\Rey\gtrsim100$. Hence the transverse force maintains a significant nonzero value throughout the range $100\lesssim\Rey\lesssim\Rey^{SS}$. For low enough separations, the force exhibits little variation with the Reynolds number in that range and even up to $\Rey=250$. For instance, $C_{\text Lu}^\text{W}$ only varies by $\pm5\%$ about a mean value close to $0.1$ for $L_\text{R}=1.5$.  Under such conditions, no stationary bifurcation takes place, the wake structure having reached some kind of `asymptotic' state that breaks down only at much higher Reynolds number with the occurrence of unsteady effects. The case $L_\text{R}=2$ represents an intermediate situation in which the transverse force is seen to increase significantly beyond $\Rey=150$, almost doubling its value at $\Rey=250$. This variation suggests that the wake structure changes significantly within this range. This was confirmed in \cite{2005_Zeng}, where it was shown that the size of the double-threaded wake structure grows dramatically from $\Rey=100$ to $\Rey=200$ (their figure 12). Since the streamwise vortices act to deflect the fluid toward the wall in the symmetry plane, the wake it more vigorously tilted in that direction as $\Rey$ increases, a trend confirmed by the comparison of panels (h) and (k) in Fig. \ref{fig:spanwiseVor}. This in turn increases the fluid velocity directed toward the fluid interior on the outer side of the streamwise vortices, hence the repelling transverse force.

\subsection{Linear shear flow}
\label{sec:5.2}
\subsubsection{Drag on a non-rotating sphere}
\label{sec:5.2.1}
\begin{figure}
	\centerline{\includegraphics[scale=0.9]{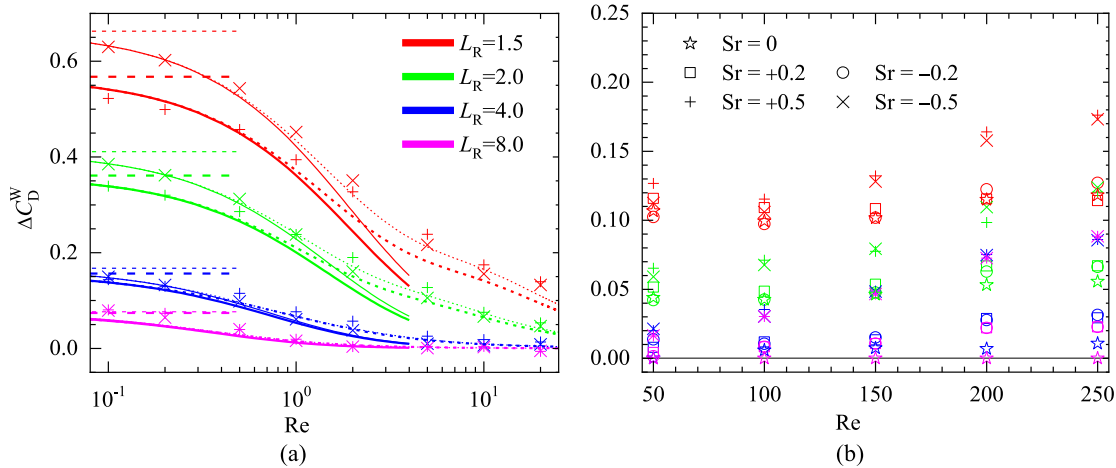}}
	\caption{Relative near-wall drag increase $\Delta C_\text D^\text W$ for a non-rotating sphere translating parallel to a wall in a linear shear flow. (a): low-to-moderate Reynolds number regime for $\Sry=\pm0.5$; (b): moderate-to-high Reynolds number regime for $\Sry=\pm0.2$ and $\pm0.5$. Symbols in (a): numerical results for $\Sry=0.5$ ($+$) and $\Sry=-0.5$ ($\times$). Dashed lines: asymptotic prediction \eqref{eq:CDWin_tf} corresponding to conditions $L_u\ll1,\,L_\omega\ll1$; solid lines: low-but-finite-$\Rey$ semiempirical expression (\ref{drag_fin}); dotted lines: low-to-moderate-$\Rey$ semiempirical expression \eqref{eq:Wshear-drag-Re1}. Thick (thin) lines correspond to positive (negative) $\Sry$.}
	\label{fig:drag_sr05_low-to-mo-Re}
\end{figure}

The drag change ratio $\Delta C_{\text D}^\text W=(C_{\text D}^\text W-C_{\text D0}^\text U)/C_{\text D0}^\text U$ is reported in Fig. \ref{fig:drag_sr05_low-to-mo-Re} for different separation distances and dimensionless shear rates. 
Let us first consider results obtained in the low-to-moderate Reynolds number regime ($0.1\leq \Rey \leq 20$) with a dimensionless shear rate $\Sry=\pm0.5$ (Fig. \ref{fig:drag_sr05_low-to-mo-Re}(a)). For $L_\text{R}\leq4$, the shear-induced drag modification is negligibly small compared with that resulting from the presence of the wall. In contrast, for the smallest two separations, the shear is found to increase (decrease) the drag when the sphere leads (lags) the fluid, which is supported by the qualitative discussion at the end of Sec. \ref{sec:4.1}. The asymptotic prediction \eqref{eq:CDWin_tf}, which is valid in the low-$\Rey$ limit provided the wall stands in the inner region of the disturbance, is in good agreement with numerical results at $\Rey=0.1$. Compared with the un-sheared case, the corresponding relative variation of $\Delta C_{\text D}^\text W$ is approximately $8\%$ for $\Sry=\pm0.5$. 
No explicit theoretical solution for $\Delta C_{\text D}^\text W$ is available for $\Sry\neq0$ when the wall stands in the outer region of the disturbance. However the relative influence of the shear is always small under the conditions considered here, and the decrease of $\Delta C_{\text D}^\text W$ up to $\Rey = 1$ is satisfactorily captured by (\ref{drag_fin}), as the solid lines in Fig. \ref{fig:drag_sr05_low-to-mo-Re} show. To extend this estimate to Reynolds numbers of $\mathcal{O}(10)$, we merely duplicate the arguments that led to \eqref{eq:Wu-drag-low-to-mo-Re} in the un-sheared case. The empirical counterpart of \eqref{eq:Wu-drag-low-to-mo-Re} is thus
\begin{equation}
\Delta C_\text D^\text W [\Rey\lesssim10] \approx  f_\text D^{\prime}(L_u) [a^\text W(\Rey,L_\text{R})]^2 \Delta C_{\text D}^\text{W-in}(L_\text{R}, \Sry)\,,
\label{eq:Wshear-drag-Re1}
\end{equation}
with $\Delta C_{\text D}^\text{W-in}(L_\text{R},\Sry)$, $f_\text D^\prime(L_u)$, and $a^\text{W}(\Rey,L_\text{R})$ as given in (\ref{eq:CDWin_tf}), (\ref{eq:fD_prime}), and \eqref{eq:aW}, respectively. The dotted lines in Fig. \ref{fig:drag_sr05_low-to-mo-Re} confirm that the corresponding predictions properly reproduce the variations of the numerical data up to $\Rey\approx10$.\\
\begin{figure}
	\centerline{\includegraphics[scale=0.85]{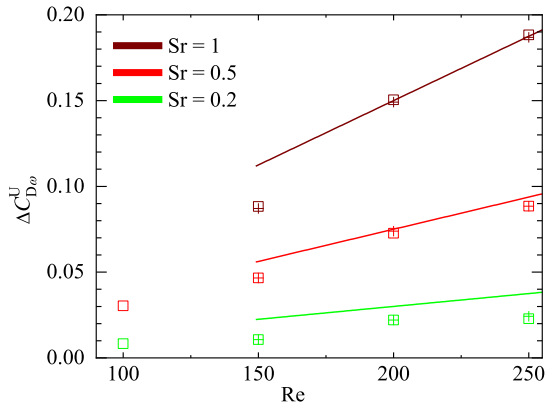}}
	\caption{Relative drag variation $\Delta C_{\text D\omega}^\text U$ on a sphere translating at moderate-to-high Reynolds number in an unbounded shear flow, with respect to the drag in a uniform stream. Symbols: numerical results for $\Sry>0$ ($\square$), and $\Sry<0$ (+). Lines: prediction of \eqref{eq:U-drag-high-Re}.}
	\label{fig:drag-ub}
\end{figure}
\indent Figure \ref{fig:drag_sr05_low-to-mo-Re}(b) displays the drag force computed for $\Rey\geq50$. While the drag still decreases with the Reynolds number up to $\Rey=100$, a systematic increase is observed at higher $\Rey$ whatever the distance to the wall. Moreover, in the same high-$\Rey$ regime, drag variations are found to be virtually independent of the sign of $\Sry$. However, for $\Sry=\mathcal{O}(1)$, the magnitude of the shear, \text{i.e.} the norm of $\Sry$, has a significant influence on the drag, with for instance a $45\%$ increase of $\Delta C_\text D^\text W$ at $\Rey=250$ from $\Sry=\pm0.2$ to $\pm0.5$. To better analyze these results, it is appropriate to consider the unbounded configuration first, in order to examine the relative drag change $\Delta C_{\text D\omega}^\text U(\Rey,\,\Sry)=\left(C_{\text D\omega}^\text U(\Rey,\,\Sry)-C_{\text D0}^\text U(\Rey)\right)/C_{\text D0}^\text U(\Rey)$ due solely to the influence of the ambient shear. Figure \ref{fig:drag-ub} shows how $\Delta C_{\text D\omega}^\text U(\Rey,\,\Sry)$ varies with both the Reynolds number and the dimensionless shear rate. An obvious symmetry argument indicates that $\Delta C_{\text D\omega}^\text U(\Rey,\,\Sry)$ cannot depend on the sign of $\Sry$, a constraint confirmed by numerical results. To magnify the shear-induced drag variations, specific runs were carried out for $\Sry = 1$. While only marginal for $|\Sry|=0.2$, the relative shear-induced drag increase is found to reach approximately $20\%$ at $\Rey=250$ with $|\Sry|=1$. Within the considered ranges of $\Sry$ and $\Rey$, $\Delta C_{\text D\omega}^\text U$ depends almost linearly on $\Sry$ and $\Rey$ beyond $\Rey \approx 150$, in agreement with the tendency observed in \cite{1999_Kurose}. 
Fitting the results obtained at $\Rey = 250$ yields
\begin{equation}
\Delta C_{\text D\omega}^\text U [\Rey=\mathcal{O}(100)] \approx 7.5\times10^{-4} |\Sry| \Rey\,,
\label{eq:U-drag-high-Re}
\end{equation}
an estimate seen to properly capture the dominant trend revealed by numerical results for $\Rey\gtrsim200$, although it over-estimates the drag increase at lower Reynolds numbers. A quantitative comparison with the findings reported in \cite{1999_Kurose} reveals that present values for $\Delta C_{\text D\omega}^\text U$ are typically twice as large. We investigated the possible origin of such a large difference, suspecting in particular that results may be contaminated by artificial confinement effects induced by the outer boundary of the computational domain, especially in the wake region.  To check this possibility, we re-ran the simulations corresponding to $\Rey=200$ on a domain twice as large as the standard one, \text{i.e.} with the outer boundary located $40d$ from the sphere center instead of $20d$, the resolution being kept unchanged close to the sphere, especially within the boundary layer, by increasing the number of nodes. 
The drag was found to vary by less than $0.3\%$ in all cases, inducing variations of $\Delta C_{\text D\omega}^\text U$ not exceeding $2\%$. These tests make us confident that present results for the high-$\Rey$ shear-induced drag increase are robust, being especially almost independent of the position of the outer boundary of the domain. In contrast, we suspect that this issue may have affected the results reported in \cite{1999_Kurose}, as that study made use of an ellipsoidal grid extending only to $10d$ upstream and downstream of the sphere and $5d$ in the direction perpendicular to the incoming flow. 
\\
\begin{figure}
	\centerline{\includegraphics[scale=0.9]{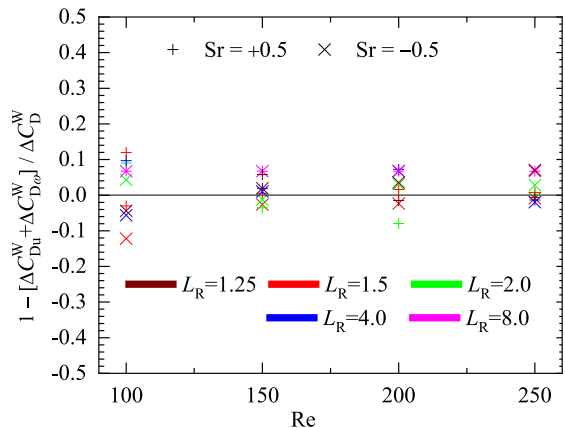}}
	\caption{Comparison for $\Sry=\pm0.5$ of the measured near-wall drag increase $\Delta C_{\text D}^\text{W}$ in the high-Reynolds-number regime with the prediction $\Delta C_{\text Du}^\text W [\Rey=\mathcal{O}(100)]+\Delta C_{\text D\omega}^\text W [\Rey=\mathcal{O}(100)]$ provided by the empirical models (\ref{eq:Wu-drag-high-Re}) and (\ref{eq:W-drag-highRe}). 
	}
	\label{fig:drag_wall-shear-interact}
\end{figure}
\begin{figure}
	\centerline{\includegraphics[scale=0.9]{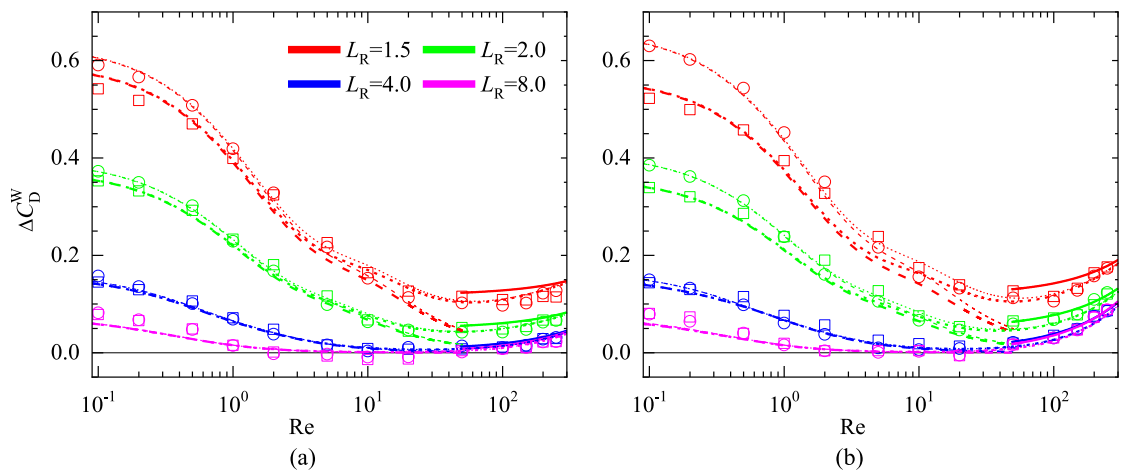}}
	\caption{Relative near-wall drag increase $\Delta C_\text D^\text W(\Rey,\Sry,L_\text{R})$ for a non-rotating sphere translating parallel to a wall in a linear shear flow, throughout the $\Rey$-range investigated numerically. (a) $\Sry=\pm0.2$; (b) $\Sry=\pm0.5$. $\square$ and $\ocircle$: numerical data for $\Sry>0$ and $\Sry<0$, respectively. Thick solid lines: high-$\Rey$ model based on the sum of (\ref{eq:Wu-drag-high-Re}) and (\ref{eq:W-drag-highRe}); thick (thin) dashed lines: low-to-moderate-$\Rey$ expression \eqref{eq:Wshear-drag-Re1} for positive (negative) $\Sry$; thick (thin) dotted lines: composite fit (\ref{eq:W-drag}) for positive (negative) $\Sry$.}
	\label{fig:drag_low-to-high-Re}
\end{figure}
\indent Coming back to the near-wall situation, we first evaluated how the observed drag variation, $\Delta C_{\text D}^\text{W}$, compares with the sum of the slip effect in the shearless case, $\Delta C_{\text Du}^\text{W}$, and the shear effect in the unbounded configuration, $\Delta C_{\text D\omega}^\text U$, as given by (\ref{eq:Wu-drag-high-Re}) and (\ref{eq:U-drag-high-Re}), respectively. It turned out that this sum consistently over-estimates $\Delta C_{\text D}^\text{W}$, and the shorter the wall-particle separation the larger the over-estimate. This finding implies that the shear-induced drag correction observed in the unbounded case is actually reduced by the presence of the wall, owing to the modifications the latter imposes on the wake structure. Keeping $\Rey$ and $\Sry$ fixed and varying $L_\text{R}$, we observed that the excess quantity $\Delta C_{\text Du}^\text{W}+\Delta C_{\text D\omega}^\text U-\Delta C_{\text D}^\text{W}$ varies as the inverse of the separation. Introducing the shear-induced drag modification in the presence of the wall, $\Delta C_{\text D\omega}^\text W$, such that $\Delta C_{\text D}^\text{W}=\Delta C_{\text Du}^\text{W}+\Delta C_{\text D\omega}^\text W$, and fitting the numerical data yields 
 \begin{equation}
\Delta C_{\text D\omega}^\text W [\Rey=\mathcal{O}(100)] \approx \left(1-0.54L_\text{R}^{-1}\right)\Delta C_{\text D\omega}^\text U [\Rey=\mathcal{O}(100)]\,,
\label{eq:W-drag-highRe}
\end{equation}
with $\Delta C_{\text D\omega}^\text U [\Rey=\mathcal{O}(100)]$ as given by (\ref{eq:U-drag-high-Re}).\\
\indent The relative difference between the observed drag variation $\Delta C_{\text D}^\text{W}$ and the prediction corresponding to the sum of (\ref{eq:Wu-drag-high-Re}) and (\ref{eq:W-drag-highRe}) is shown in Fig. \ref{fig:drag_wall-shear-interact} for $\Sry=\pm0.5$. 
It is seen that, beyond $\Rey\approx100$, this difference never exceeds $7\%$, confirming that the above empirical model properly captures the near-wall drag variations in the high-$\Rey$ regime.
Numerical results obtained throughout the range $0.1\leq \Rey\leq250$, together with the corresponding predictions based on the semiempirical expressions established above, are displayed in Fig. \ref{fig:drag_low-to-high-Re}. 
In a way similar to \eqref{eq:Wu-drag}, a purely empirical expression combining linearly the models previously established in the low-to-moderate Reynolds number regime [Eq. \eqref{eq:Wshear-drag-Re1}] and high-Reynolds number regime [Eqs. \eqref{eq:Wu-drag-high-Re}, \eqref{eq:U-drag-high-Re} and \eqref{eq:W-drag-highRe}] may be designed to improve the estimate of the drag increase in the intermediate range $10\leq \Rey\leq100$. As the dotted lines in Fig. \ref{fig:drag_low-to-high-Re} show, the composite expression 
 \begin{equation}
\Delta C_\text D^\text W \approx \Delta C_\text D^\text W [\Rey=\mathcal{O}(1-10)]+
c_{D\omega\infty}\left\{\Delta C_{\text Du}^\text W[\Rey=\mathcal{O}(100)]+\Delta C_{\text D\omega}^\text W[\Rey=\mathcal{O}(100)]\right\}\quad\mbox{with}\quad  c_{D\omega\infty}=1-\exp^{-0.035\Rey^{0.75}}\,
\label{eq:W-drag}
\end{equation}
correctly fits the numerical data throughout the entire range of Reynolds number.

\subsubsection{Lift on a non-rotating sphere}\label{sec:5.2.2}	

\begin{figure}
	\centerline{\includegraphics[scale=0.9]{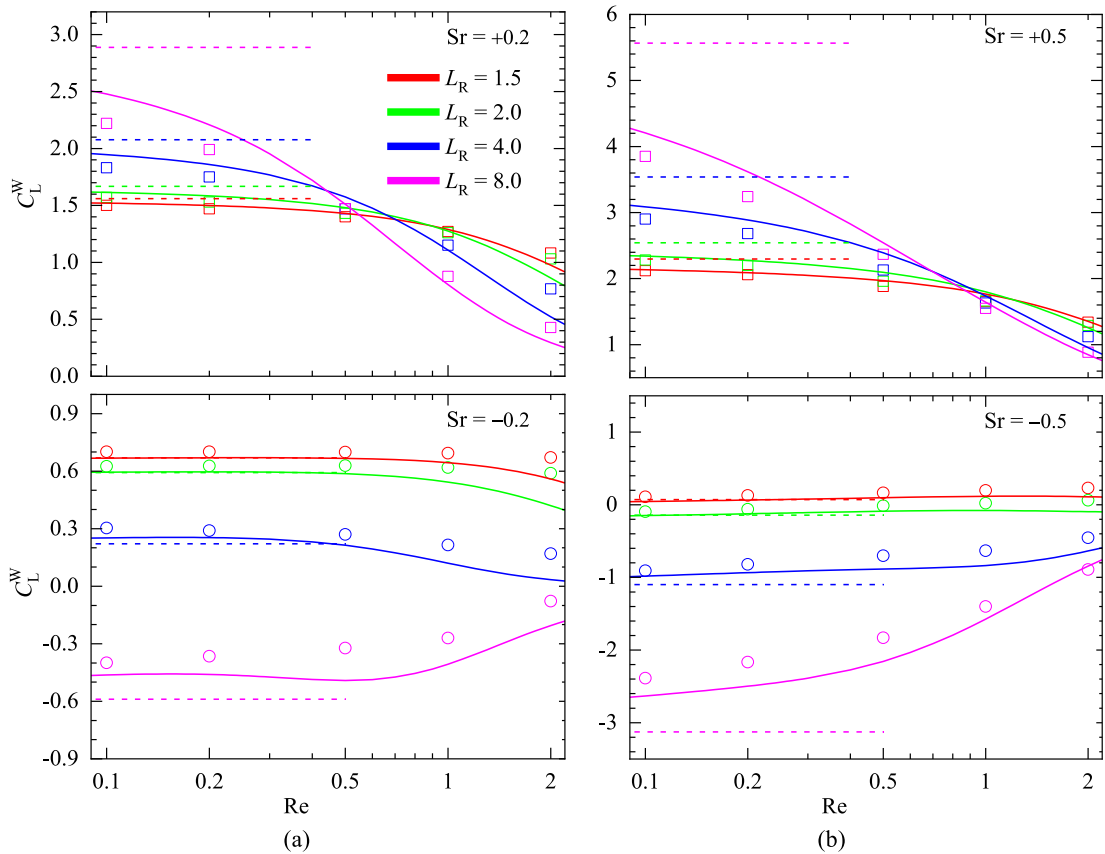}}
	\caption{Variations of the lift coefficient $C_\text L^\text W(\Rey,\Sry,L_\text{R})$ for a non-rotating sphere translating parallel to a wall in a linear shear flow at $\Rey\leq2$. (a) $\Sry=\pm0.2$; (b) $\Sry=\pm0.5$. $\square$ and $\ocircle$: numerical data for $\Sry>0$ and $\Sry<0$, respectively. Dashed lines: inner solution \eqref{eq:CLWin} corresponding to conditions $L_u\ll1,\,L_\omega\ll1$; solid lines:  finite-$\Rey$ expression (\ref{eq:CWlowRe}) with $f_\text L$ and $h_\text L$ as given by (\ref{eq:fhL0}a) and (\ref{eq:fhL0}b), respectively.}
	\label{fig:lift_low-Re}
\end{figure}

Figure \ref{fig:lift_low-Re} shows the computed lift coefficient $C_{\text L}^\text W(\Rey,\Sry,L_\text{R})$ up to $\Rey=2$ for various normalized shear rates and separation distances. Variations of $C_{\text L}^\text W$ with increasing $\Rey$ and $L_\text{R}$, as well as the form of the interplay between the shear- and slip-induced contributions, are consistent with those observed with clean spherical bubbles in \cite{2020_Shi_a}. For this reason, the reader is referred to section 6.2.2 of that reference for a discussion of the physical mechanisms governing the variations of the lift force with $\Rey$, $\Sry$ and $L_\text{R}$ revealed by Fig. \ref{fig:lift_low-Re}. The reason why the lift force acting on a rigid sphere or a clean spherical bubble behave similarly in this regime has been established in \cite{1997_Legendre} and extended to near-wall configurations in \cite{2002_Takemura,2003_Magnaudet}. Specifically, these analyses indicate that, to leading order, shear-, wall-, and the combined lift forces acting on a rigid sphere in the low-but-finite Reynolds number regime differ from those on a clean spherical bubble only by a pre-factor of $(3/2)^2$, $3/2$ being the strength ratio of the respective Stokeslets. That the lift force on a rigid sphere at a given $\Sry$ and $L_\text{R}$ is larger than that on a clean bubble by a factor  of $(3/2)^2=2.25$ may be confirmed by comparing present data at $\Rey = 0.1$ with their counterparts in figure 18 of \cite{2020_Shi_a}. 
As expected, the dashed lines in Fig. \ref{fig:lift_low-Re} indicate that the asymptotic expression \eqref{eq:CLWin} corresponding to situations in which the wall stands in the inner region of the disturbance predicts the computed lift force well for small enough $L_\text{R}$ and $\Rey$, typically  $L_\text{R}<4$ and $\Rey\leq0.2$ for both shear rates. When $L_\text{R}$ or $\text{Re}$ increases, the wall shifts to the outer region of the disturbance and \eqref{eq:CLWin} fails to capture the variations of the lift coefficient. In contrast, the finite-$\Rey$ expression (\ref{eq:CWlowRe}) (along with (\ref{eq:fhL0}a) and (\ref{eq:fhL0}b) for $f_\text L$ and $h_\text L$, respectively) 
(solid lines in Fig. \ref{fig:lift_low-Re}) properly accounts for these effects up to $\Rey = 2$.
\begin{figure}
	\centerline{\includegraphics[scale=0.9]{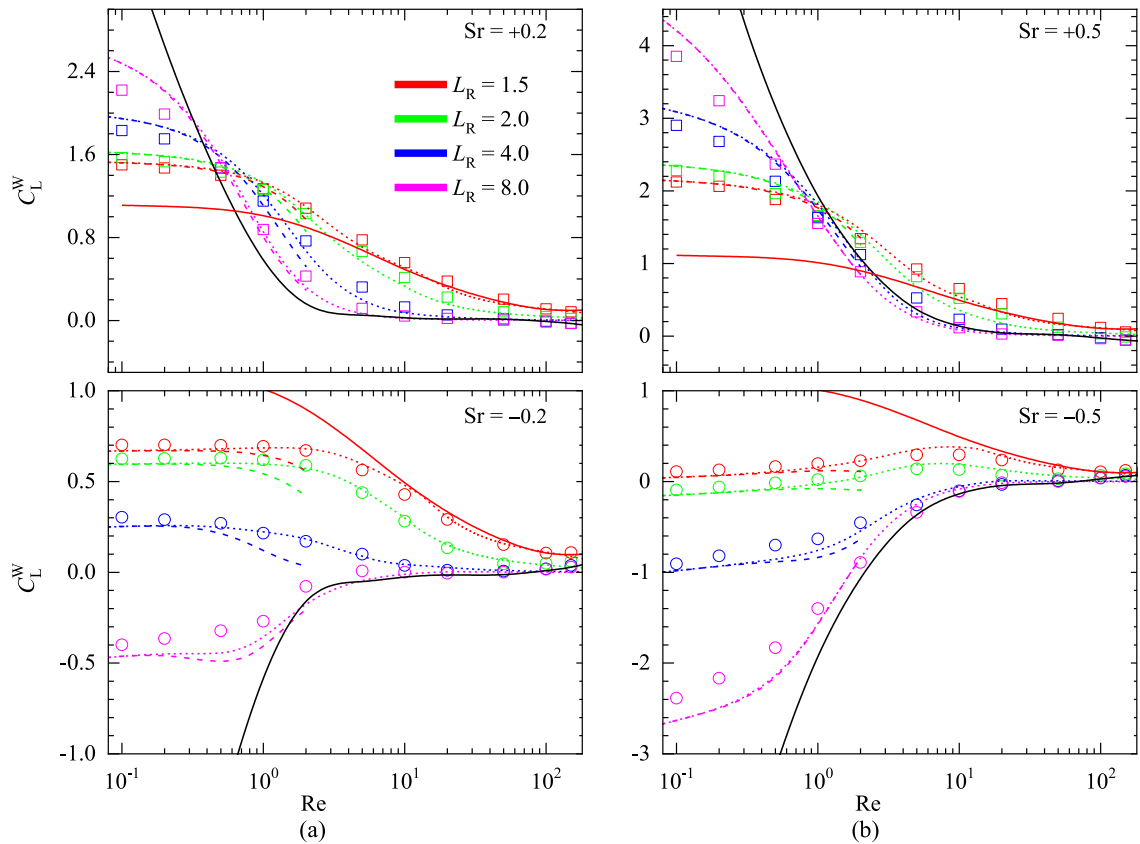}}
	\caption{Lift coefficient $C_\text L^\text W(\Rey,\Sry,L_\text{R})$ for a non-rotating sphere translating parallel to a wall in a linear shear flow at $\Rey\leq150$. (a) $\Sry=\pm0.2$; (b) $\Sry=\pm0.5$. $\square$ and $\ocircle$: numerical data for $\Sry>0$ and $\Sry<0$, respectively. Dashed lines: finite-$\Rey$ prediction (\ref{eq:CWlowRe}); dotted lines: prediction \eqref{eq:CLWmoderRe}; black solid lines: lift coefficient $C_{\text L\omega}^\text U$ in an unbounded shear flow; red solid lines: lift coefficient $C_{\text Lu}^\text W$ at $L_\text{R}=1.5$ in a wall-bounded fluid at rest.}
	\label{fig:lift_low-to-mo-Re}
\end{figure}

Figure \ref{fig:lift_low-to-mo-Re} summarizes the behavior of $C_\text L^\text W(\Rey,\Sry,L_\text{R})$ for $0.1\leq \Rey\leq150$ at various normalized shear rates and separation distances. The lift force is seen to vary sharply with both the Reynolds number and the separation distance for $\Rey\lesssim10$, while at higher Reynolds number substantial variations only subsist for $L_\text{R}\lesssim2$. For larger separations, the magnitude of the lift force gets close to that found in an unbounded flow (black solid lines in Fig. \ref{fig:lift_low-to-mo-Re}). Indeed, at such Reynolds numbers the thickness of the boundary layer around the sphere is small enough for the vortical interaction with the wall to have only a secondary influence on the lift force (see Fig. \ref{fig:spanwiseVor}). Effects caused by the shear may be qualitatively estimated by comparing $C_\text L^\text W(\Rey,\Sry,L_\text{R})$ with its counterpart in the shearless situation, $C_{\text Lu}^\text W(\Rey,L_\text{R})$,  shown in Fig. \ref{fig:lift_low-to-mo-Re} for $L_\text{R}=1.5$ (red solid lines). Clearly, the slip effect dominates for $\Rey\gtrsim10$.  Influence of the shear becomes more pronounced or even dominant at lower Reynolds numbers. For large separations, it increases (decreases) the total lift force well beyond (below) the level reached in the shearless case for $\Sry>0$ ($\Sry<0$). 
The influence of $\Sry$ weakens as $L_\text{R}$ decreases, the presence of the wall inhibiting the development of the wake. Selecting for instance $L_\text{R} = 1.5$, $\Rey = 1$, and $\Sry= 0.5$, the difference between $C_\text L^\text W$ and  $C_{\text Lu}^\text W$ is $0.66$, three times less than the lift coefficient $C_{\text L\omega}^\text U=1.93$ in the unbounded case. 
Based on the above observations, and disregarding the small shear-induced contribution to the lift beyond $\Rey\approx10$, the finite-$\Rey$ expression (\ref{eq:CWlowRe}) may be extended to moderate Reynolds numbers in the form \\
\begin{eqnarray}
\label{eq:CLWmoderRe}
C_\text L^\text W[\Rey=\mathcal{O}(1-100)] &\approx& g_\text L C_{\text Lu}^\text W [\Rey=\mathcal{O}(1-100)]+h_\text LC_{\text L\omega}^\text{U}(\Rey\ll1)\,,\\
\label{gl}
\mbox{with}\quad g_\text L(L_\omega,\varepsilon,\Rey) &=&\exp^{-0.22\varepsilon^{0.8}L_\omega^{2.5}\text{exp}(-0.01\Rey^2)}\,,
\\\nonumber
\end{eqnarray}
and $C_{\text Lu}^\text W [\Rey=\mathcal{O}(1-100)]$, $h_\text L$ and $C_{\text L\omega}^\text{U}(\Rey\ll1)$ as provided by \eqref{eq:CLuW_moderRe}, (\ref{eq:fhL0}b) and \eqref{eq:U_low_Re}, respectively. Note that, similar to expressions \eqref{eq:CWlowRe} and (\ref{eq:fhL0}a) in the low-but-finite Reynolds number regime, \eqref{eq:CLWmoderRe} and \eqref{gl} indicate that the wall and shear effects do not superimpose linearly, as the pre-factor $g_\text{L}$ for the former involves $\Sry$ through the presence of $\varepsilon$ and $L_\omega$. 
As shown in Fig. \ref{fig:lift_low-to-mo-Re}, \eqref{eq:CLWmoderRe} fits all numerical predictions well throughout the range $0.1\leq \Rey\leq100$ for $L_\text{R}\leq8$. \vspace{1mm}\\
\begin{figure}
	\centerline{\includegraphics[scale=0.65]{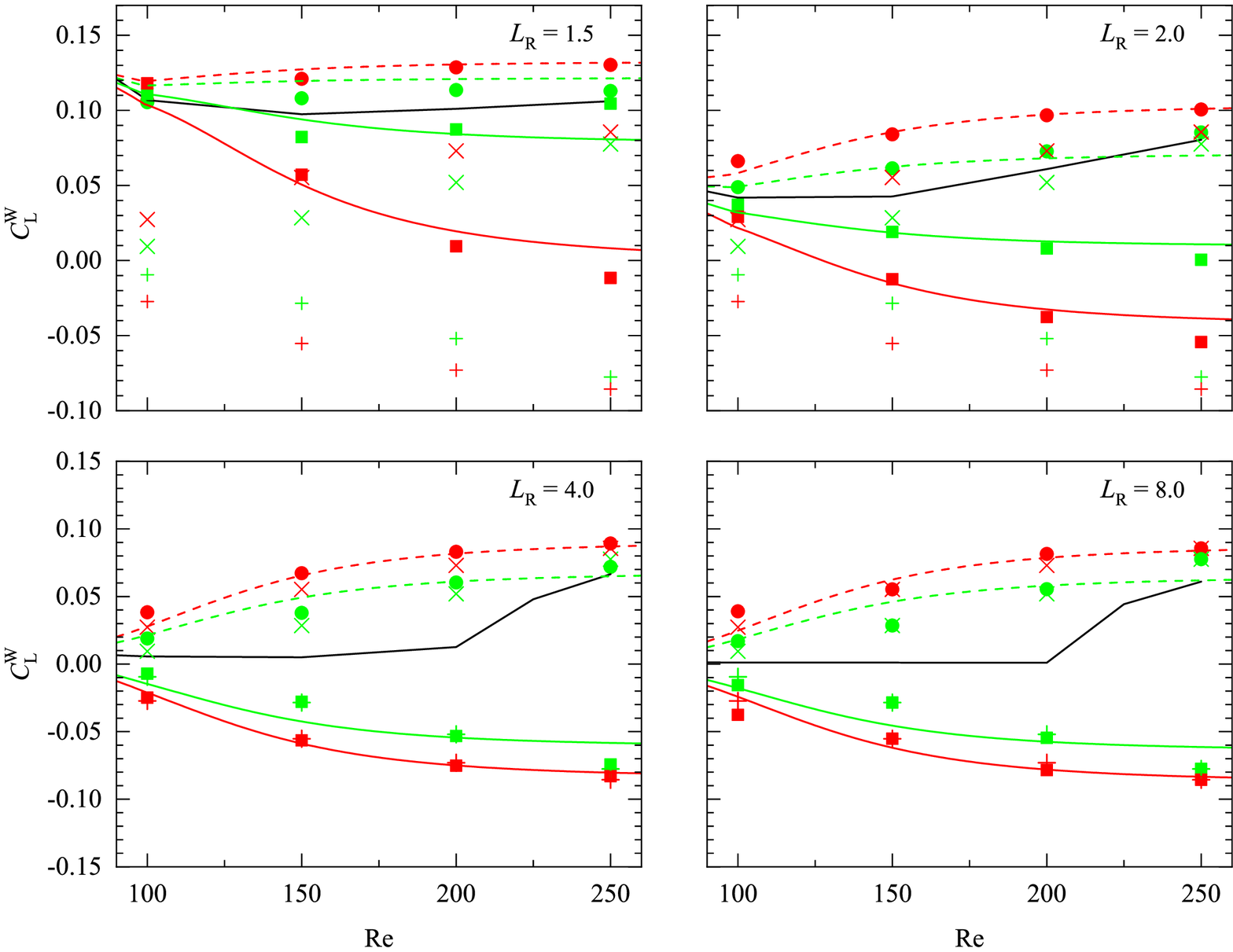}}
	\caption{Variations of the lift coefficient $C_\text L^\text W(\Rey,\Sry,L_\text{R})$ in the range $100\leq\Rey\leq250$ for a non-rotating sphere translating parallel to a wall in a linear shear flow. ${\color{green}{\blacksquare}}: \Sry=+0.2$, ${\color{green}{\medbullet}}: \Sry=-0.2$, ${\color{red}{\blacksquare}}: \Sry=+0.5$, ${\color{red}{\medbullet}}: \Sry=-0.5$. 
	${\color{red}+}$ (${\color{green}+}$): $\Sry=+0.5$ ($+0.2$) in the unbounded configuration; same with ${\color{red}\times}$ (${\color{green}\times}$) for $\Sry=-0.5$ ($-0.2$). Black solid line: $\Sry=0$; green solid (dashed) line: prediction of \eqref{eq:lift_highRe} for $\Sry=+0.2$ ($-0.2$); same with the red solid (dashed) line for $\Sry=+0.5$ ($-0.5$).}
	\label{fig:lift_high-Re}
\end{figure}
\indent We now turn to the high-Reynolds-number regime, say $\Rey\gtrsim100$. At such Reynolds numbers, the shear-induced lift force observed in the unbounded case has changed sign, as illustrated in Fig. \ref{fig:streamwise_Vor_ub}. For small separations, typically $L_\text{R}\lesssim2$, the slip-induced transverse force discussed in Sec. \ref{sec:5.1.2} remains non-negligible up to $\Rey = 200$. When the two effects combine, the streamwise vorticity distribution illustrated in Fig. \ref{fig:streamwise_Vor_wb} reveals that the two mechanisms act in an antagonistic (cooperative) manner when $\Sry$ is positive (negative). This is confirmed in Fig. \ref{fig:lift_high-Re}, where, taking the results corresponding to $\Sry = 0$ (black line) as reference, a negative $\Sry$ is seen to contribute positively to the lift force and \textit{vice versa}, unlike the low-to-moderate Reynolds number phenomenology. Moreover, slip- and shear-induced effects combine in a strongly nonlinear manner. Indeed, for a given magnitude of $\Sry$ and a decreasing $L_\text{R}$, the shear-induced variation, $|C_\text L^\text W-C_{\text Lu}^\text W|$, is seen to become significantly larger when $\Sry$ is positive (compare in particular the data pairs corresponding to $\Sry = \pm0.5$ and $L_\text{R}=1.5$).  For $\Rey\gtrsim 200$ and $\Sry=0$, the stationary imperfect bifurcation discussed in Sec. \ref{sec:4.1} takes place when the separation is large enough, causing a sharp increase in the transverse force, as the black lines in Fig. \ref{fig:lift_high-Re} confirm for $L_\text{R}\geq4$. For smaller separations, or for arbitrary separations in the presence of shear, no bifurcation takes place in the wake in this $\Rey$-range, since the flow past the sphere is fully three-dimensional whatever $\Rey$. This is the reason why the increase in the magnitude of $C_\text L^\text W$ with $\Rey$ is much more gradual in such situations. As the comparison with data corresponding to the unbounded sheared configuration (crosses) reveals, the wall no longer influences the lift force for $L_\text{R}\gtrsim4$. Conversely, for smaller separations, the lift force found for $\Sry>0$ ($\Sry<0$) reduces (increases) gradually compared to its value in an unbounded flow as $L_\text{R}$ decreases. \\
\indent We sought a correlation capable of reproducing the above trends. For this purpose, since the shear-induced lift in an unbounded flow changes sign for $\Rey\approx50$,  we used the expression provided in \eqref{eq:lift_loth} for $C_{\text L\omega}^\text{U}[\Rey=\mathcal{O}(100)]$. Then, Fig. \ref{fig:lift_sr0_low-and-high-Re}(b) suggests that the slip-induced contribution $C_{\text Lu}^\text W$ is almost constant beyond $\Rey=100$ when no stationary bifurcation takes place in the wake (see the data set corresponding to $L_\text{R}=1.5$). This situation also holds when $\Sry\neq0$, since the wake is three-dimensional whatever $\Rey$, similar to the configuration $\Sry=0$ when  $L_\text{R}$ is small. Therefore it sounds reasonable to assume that, at a given separation, $C_{\text Lu}^\text W(\Sry\neq0,\Rey\geq100)$ is close to $C_{\text Lu}^\text W(\Rey=100)$ for $\Sry=0$, as provided by \eqref{eq:CLuW_moderRe} for $\Rey=100$. Last, $C_{\text L\omega}^\text{U}[\Rey=\mathcal{O}(100)]$ has to be weighted by a pre-factor $k_\text{L}(L_\text{R},\Rey)$, in order to mimic the increasingly asymmetric magnitude of the lift force according to the sign of $\Sry$ when $L_\text{R}$ becomes small. We finally obtained
\begin{eqnarray}
\label{eq:lift_highRe}
C_\text L^\text W[\Rey=\mathcal{O}(100)] &\approx& C_{\text Lu}^\text W (\Rey=100)+ k_\text{L}C_{\text L\omega}^\text{U}[\Rey=\mathcal{O}(100)]\,,\\
\label{eq:coeffhighRe}
\mbox{with}\quad  k_\text{L}(L_\text{R},\Rey)&=&1- \exp(-0.034 L_\text{R}^{6}|\Sry|^{0.75}) + (1+\sgn(\Sry))\exp(-0.048 L_\text{R}^{4.5}|\Sry|^{-1})\exp^{-(0.009 \Rey)^{-4}}  \,,                                                            
\end{eqnarray}
with $C_{\text Lu}^\text W (\Rey=100)$ and $C_{\text L\omega}^\text{U}[\Rey=\mathcal{O}(100)]$ as provided by \eqref{eq:CLuW_moderRe} and \eqref{eq:lift_loth}, respectively. The solid and dashed lines in Fig. \ref{fig:lift_high-Re} confirm that this correlation properly captures the dramatic variations induced by the wall on the lift force, including the asymmetry observed between negative and positive relative shear rates.


\subsubsection{Effects of sphere rotation}\label{sec:5.2.3}	


\begin{figure}
	\centerline{\includegraphics[scale=0.7]{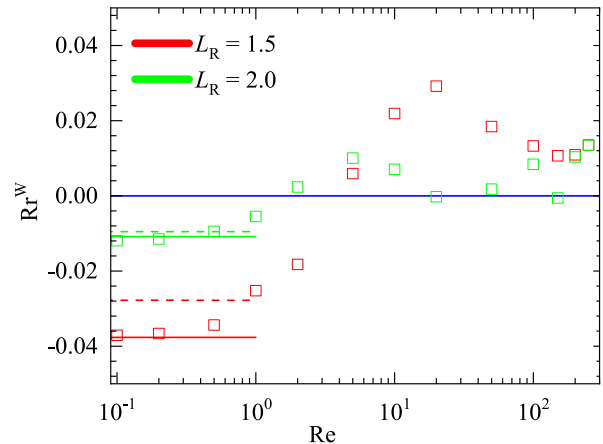}}
	\caption{Variations with $\Rey$ and $L_\text{R}$ of the rotation rate $\Rry^\text{W}$ of a torque-free sphere translating parallel to a wall in a fluid at rest. Symbols: numerical results; dashed lines: asymptotic prediction \eqref{eq:spin_rate_w} corresponding to the condition $L_u\ll1$; solid lines: empirical extension  (\ref{eq:spin_rate_w0}) of \eqref{eq:spin_rate_w} based on the exact zero-$\Rey$ prediction of \citep{1964_ONeill,1967_Goldman_a}.}
	\label{fig:spin-rate_sr0}
\end{figure}

\begin{figure}
	\centerline{\includegraphics[scale=0.65]{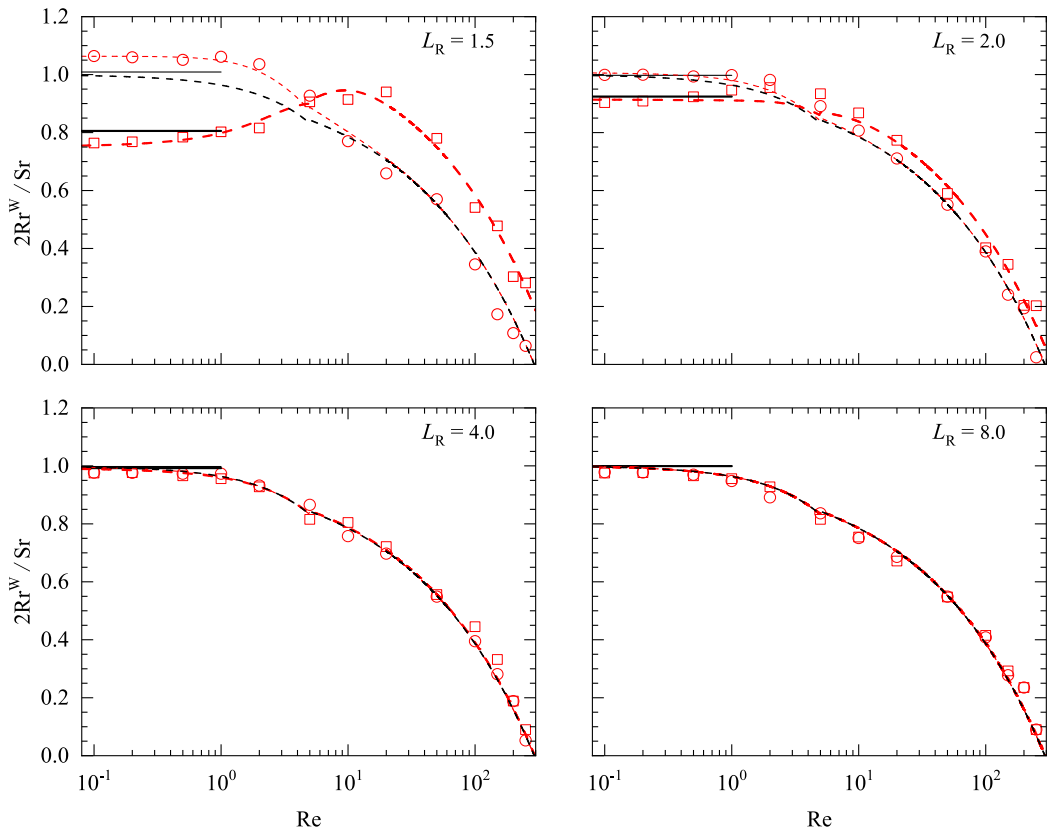}}
	\caption{Variations with $\Rey$ and $L_\text{R}$ of the normalized rotation rate $2\Rry^\text{W}/\Sry$ of a torque-free sphere translating parallel to a wall in a linear shear flow with $|\Sry|=0.5$. {\color{red}{$\square$}}  and  {\color{red}{$\ocircle$}}: numerical data for $\Sry>0$ and $\Sry<0$, respectively. Solid lines: inner solution \eqref{eq:spin_rate_w} corresponding to conditions $L_u\ll1,\,L_\omega\ll1$; black dashed lines: empirical prediction \eqref{eq:spin_rate_ua}-\eqref{eq:spin_rate_ub} from \cite{2002_Bagchi} in an unbounded shear flow; red dashed lines: empirical fit \eqref{eq:spin_rate_w2}. Thick and thin lines refer to predictions for $\Sry=+0.5$ and $\Sry=-0.5$, respectively.}
	\label{fig:spin-rate}
\end{figure}

The rotation rate of a torque-free sphere translating parallel to the wall in a fluid at rest is shown in Fig. \ref{fig:spin-rate_sr0} for the shortest two separations, $L_\text{R}=1.5$ and $\,2$, over the range $0.1\leq \Rey\leq250$. $\Rry^\text{W}$ is seen to change from negative at low Reynolds number (in agreement with Faxén's prediction \eqref{eq:spin_rate_w} \cite{1965_Happel}) to positive beyond a critical $\mathcal{O}(1)$-Reynolds number, $\Rey_{c_T}$. This critical value depends on $L_\text{R}$, and numerical results indicate $\Rey_{c_T}\approx4$ for $L_\text{R}=1.5$ and $\Rey_{c_T}\approx2$ for $L_\text{R}=2$. The low-$\Rey$ numerical values are found to exceed those predicted by \eqref{eq:spin_rate_w}, especially for $L_\text{R}=1.5$. We compared the exact creeping-flow values provided in \cite{1967_Goldman_a} (recomputed from the exact solution of \cite{1964_ONeill}) with Faxén's approximate prediction \eqref{eq:spin_rate_w} and found that the latter significantly under-estimates the former for $L_\text{R}\lesssim2$. A significantly better approximation, reproducing the exact prediction down to $L_\text{R}\approx1.1$, is provided by the semiempirical extension of \eqref{eq:spin_rate_w}
\begin{equation}
\Rry^\text{W}(L_\text{R}, \Rey\ll1)\approx- \frac{3}{16}L_\text{R}^{-4}\left(1-\frac{3}{8}L_\text{R}^{-1}+0.9L_\text{R}^{-3}\right)\,,
\label{eq:spin_rate_w0}
\end{equation}
suggesting that higher-order corrections neglected in Faxén's solution are required to accurately estimate $\Rry^\text{W}$ at such small separations. As the solid lines in Fig. \ref{fig:spin-rate_sr0} indicate, numerical results at $\Rey=0.1$ are in close agreement with (\ref{eq:spin_rate_w0}). For $\Rey>\Rey_{c_T}$, the rotation rate first increases up to a maximum ($\Rry^\text{W}\approx0.01$ at $\Rey\approx5$ for $L_\text{R}=2$, $\Rry^\text{W}\approx0.029$ at $\Rey\approx20$ for $L_\text{R}=1.5$), before exhibiting non-monotonic variations with both the Reynolds number and the separation distance, especially beyond $\Rey=100$. A qualitatively similar behavior has been reported in \citep{2005_Zeng} for the torque on a non-rotating sphere in the same range of separations. We hypothesize that subtle variations in the flow structure in the sphere vicinity (see figure 3 in \citep{2005_Zeng}) are responsible for this complex behavior.\\
\indent Figure \ref{fig:spin-rate} summarizes the  normalized rotation rate $2\Rry^\text{W}/\Sry$ corresponding to the torque-free condition, as computed for $0.1\leq \Rey\leq250$ at various separation distances. Only the `strong' relative shear rate $|\Sry|=0.5$ is considered, so as to obtain sizeable effects of the sphere rotation. At low Reynolds numbers and small separations ($0.1\leq \Rey\lesssim5$, $L_\text{R}\leq2$), numerical data indicate that spheres moving under $\Sry<0$-conditions rotate faster than those moving with $\Sry>0$. This difference is the consequence of the opposite signs of the shear-induced and slip-induced contributions to the sphere rotation in the low-Reynolds number regime, as is apparent in \eqref{eq:spin_rate_w} (in the configuration of Fig. \ref{fig:schem_bub_mov}, these two contributions yield clockwise and counter-clockwise rotations, respectively). This asymptotic prediction (solid lines in Fig. \ref{fig:spin-rate}) is in good agreement with the numerical data up to $\Rey = 0.5$. A slight under-estimate (over-estimate) is noticed when $L_\text{R}=1.5$ for $\Sry<0$ ($\Sry>0$), in line with the aforementioned under-estimate of the slip effect by \eqref{eq:spin_rate_w} at short separations.  
The influence of the sign of $\Sry$ on the magnitude of $\Rry^\text{W}$ is seen to reverse at somewhat higher $\Rey$, the rotation rate becoming larger for positive relative shear rates than for negative ones when the Reynolds number exceeds the critical value $\Rey\approx6$ ($\Rey\approx3$) for $L_\text{R}=1.5$ ($L_\text{R}=2$). This change is a direct consequence of the change of sign of the slip-induced rotation in a fluid at rest, as described above. 
Thus, when the sphere is allowed to rotate in the moderately inertial regime, the direction of the slip-induced rotation is opposite to that found in the low-$\Rey$ regime, leading to a cooperative (antagonistic) effect with the shear when $\Sry$ is positive (negative). Beyond $L_\text{R}=2$, the normalized rotation rates obtained with positive and negative $\Sry$ are virtually identical, suggesting that the slip effect has become negligible at such separations compared to that of the shear. Comparing the two panels at $L_\text{R}=4$ and $L_\text{R}=8$ indicates that the shear effect itself is barely affected by the presence of the wall at such separations, the rotation rates found at a given $\Rey$ being very close for both values of $L_\text{R}$. This conclusion is reinforced by the good agreement between present results for $L_\text{R}\geq4$ and the fit \eqref{eq:spin_rate_ua}-\eqref{eq:spin_rate_ub} provided in \cite{2002_Bagchi} (black dashed lines in Fig. \ref{fig:spin-rate}), which is based on numerical results obtained in an unbounded shear flow. Both sets of results show that the rotation rate gradually decreases as the Reynolds number increases, and is reduced to approximately $40\%$ ($20\%$) of the low-$\Rey$ value $\Rry=\frac{1}{2}\Sry$ at $\Rey=100$ ($200$). Remarkably, results at the lowest two separations reveal that the rotation rate is altered by the presence of the wall in a very dissimilar manner depending on the sign of $\Sry$ in the moderate-to-large Reynolds number regime, say $\Rey\gtrsim10$: while $\Rry^\text{W}$ is significantly larger than the rotation rate found in an unbounded shear flow when $\Sry$ is positive (even for $\Rey\gtrsim100$), the wall does not seem to have any significant effect for $\Rey\geq10$ when $\Sry$ is negative.\\
\indent To account for these various effects, we sought an empirical fit tending toward \eqref{eq:spin_rate_w} when $\Rey\rightarrow0$ (with the empirical modification introduced in \eqref{eq:spin_rate_w0}) and toward \eqref{eq:spin_rate_ua}-\eqref{eq:spin_rate_ub} when $L_\text{R}\rightarrow\infty$, while taking into account the aforementioned asymmetric $\sgn(\Sry)$-dependent influence of the wall at moderate-to-large $\Rey$. We found that the best fit satisfying these requirements is 
\begin{equation}
\Rry^\text{W} \approx - \frac{3}{16}f_\text L^{\prime}L_\text{R}^{-4}\left\{1-\frac{3}{8}L_\text{R}^{-1}+0.9L_\text{R}^{-3}\right\}+\left\{f_{\Omega}^\text U+0.5L_\text{R}^{-4}\tanh(\frac{\Rey}{2})(1+\sgn(\Sry)) \right\}\left\{1-\frac{5}{16}L_\text{R}^{-3}\exp(-0.5\Rey)\right\}\frac{\Sry}{2}\,,
\label{eq:spin_rate_w2}
\end{equation}
with $f_\text L^{\prime}$ and $f_{\Omega}^\text U$ as given in \eqref{eq:fL_prime} and \eqref{eq:spin_rate_ub}, respectively. 
As the dashed lines in Fig. \ref{fig:spin-rate} show, \eqref{eq:spin_rate_w2} satisfactorily matches the numerical data throughout the considered range of $\Rey$ and $L_\text{R}$. 
\begin{figure}
	\centerline{\includegraphics[scale=0.8]{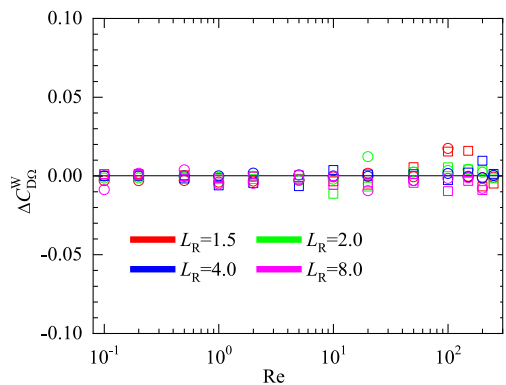}}
	\caption{Influence of the sphere rotation induced by the torque-free condition on the relative drag increase for $|\Sry|=0.5$. Symbols $\square$  and $\ocircle$ refer to $\Sry=+0.5$ and $\Sry=-0.5$, respectively.}
	\label{fig:drag_spin}
\end{figure}

The difference $\Delta C_{\text D \Omega}^\text W$ between the relative drag variations $\Delta C_{\text D}^\text W$ respectively found in the torque-free and non-rotating near-wall configurations for a given set of $(\Rey,\Sry,L_\text{R})$ is shown in Fig. \ref{fig:drag_spin}. Throughout the considered range of parameters, $\Delta C_{\text D \Omega}^\text W$ is less than $2\%$, indicating that the sphere rotation has only a marginal effect on the drag. Note that $\Delta C_{\text D \Omega}^\text W$ is even less than $1\%$ for $\Rey<100$, the largest influence of the rotation being observed in the high-Reynolds-number regime. This weak change in the drag force confirms the findings of \cite{2005_Zeng}. It is in line with the comments made in Sec. \ref{sec:4.2} regarding the tiny changes induced in the spanwise vorticity field by the sphere rotation resulting from the torque-free condition. At low Reynolds number, it is also in line with the theoretical predictions \eqref{eq:spin_rate_w} and \eqref{eq:CDWin_tf} which indicate that the drag force is affected by the particle rotation only at $\mathcal{O}(L_\text{R}^{-8})$. 
\begin{figure}
	\centerline{\includegraphics[scale=0.7]{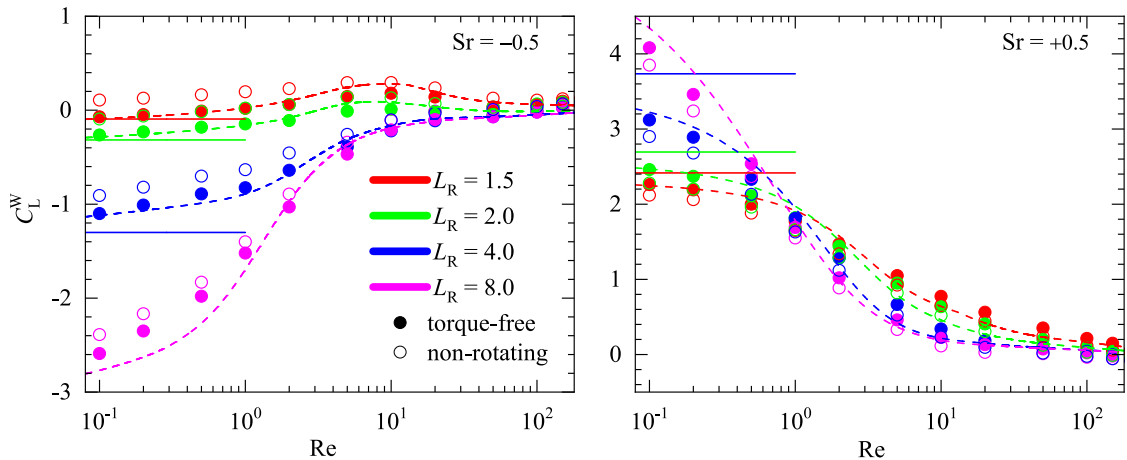}}
	\caption{Influence of the sphere rotation induced by the torque-free condition on the lift force for $|\Sry|=0.5$ and $0.1\leq \Rey\leq150$. Closed and open symbols refer to the lift coefficient for torque-free and non-rotating spheres, respectively. Solid lines: asymptotic prediction \eqref{eq:CLWin_tf} corresponding to conditions $L_u\ll1,\,L_\omega\ll1,\,L_\Omega\ll1$; dashed lines: empirical prediction \eqref{eq:lift_tf}.}
	\label{fig:lift_spin}
\end{figure}

\begin{figure}
	\centerline{\includegraphics[scale=0.65]{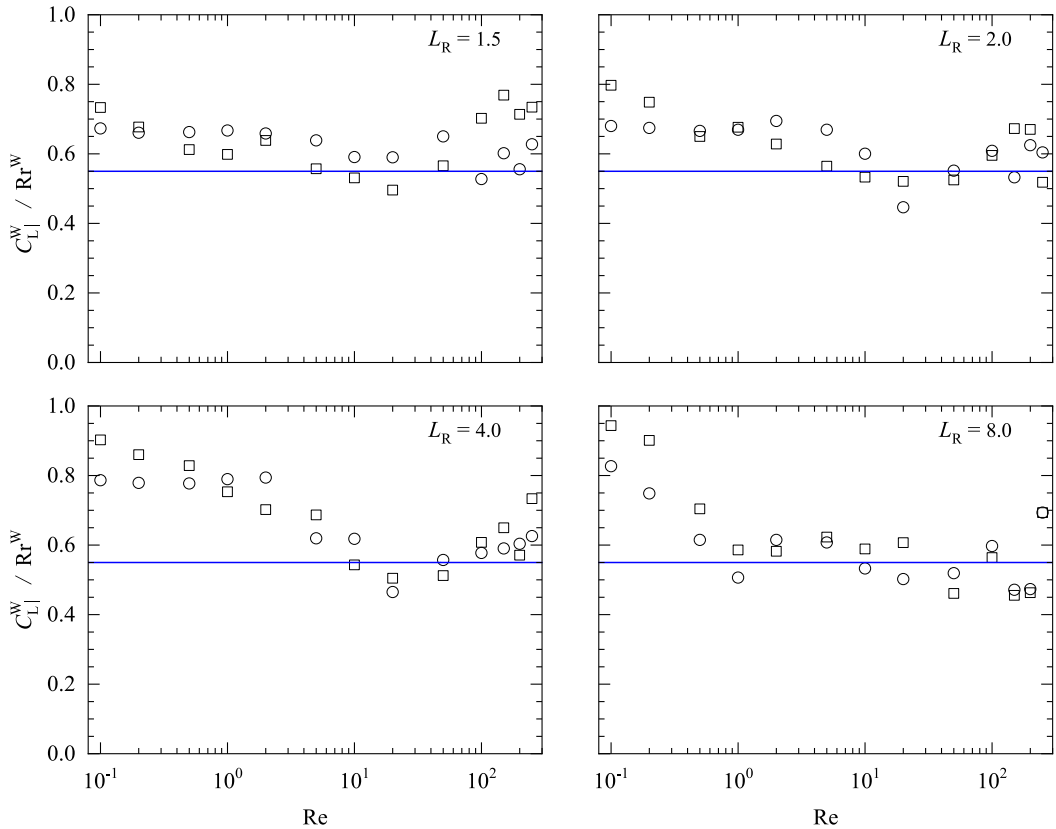}}
	\caption{Change in the lift coefficient due to the sphere rotation induced by the torque-free rotation. Values of $C_{\text L\Omega}^\text W$ are divided by the rotation rate $\Rry$ to provide a better collapse. $\square$ and $\ocircle$ symbols refer to numerical data for $\Sry>0$ and $\Sry<0$, respectively. Solid lines: approximation $C_{\text L\Omega}^\text U/\Rry \approx 0.55$ established in \citep{2002_Bagchi} in an unbounded shear flow.}
	\label{fig:lift_spin-coef}
\end{figure}

Things are somewhat different regarding the lift force. As seen in Fig. \ref{fig:lift_spin}, the lift force in the torque-free case (closed symbols) slightly but consistently differs from its counterpart in the non-rotating case (open symbols) for large enough relative shear rates (here $|\Sry|=0.5$). The rotation provides a positive contribution when the sphere lags the fluid ($\Sry>0$) and \textit{vice versa}. Whatever the sign of $\Sry$, this effect reduces as $\Rey$ increases, in a manner consistent with the variation of the torque-free rotation rate observed in Fig. \ref{fig:spin-rate}. The asymptotic prediction \eqref{eq:CLWin_tf} derived under conditions $L_u\ll1,\,L_\omega\ll1,\,L_\Omega\ll1$ is in good agreement with the data obtained at $\Rey = 0.1$ up to $L_\text{R}\approx2$ for $\Sry>0$, and $L_\text{R}\approx4$ for $\Sry<0$. That the range of accuracy of the asymptotic prediction is somewhat larger for negative relative shear rates is a property shared with the non-rotating case (compare the left and right panels in Fig. \ref{fig:lift_low-Re}). This is presumably because the nonlinear interaction between the slip-induced and shear-induced mechanisms contributing to the lift force is somewhat weaker when the two mechanisms are antagonistic, i.e. when $\Sry<0$. \\
\indent To extend empirically the validity of \eqref{eq:CLWin_tf} toward moderate Reynolds numbers, the change $C_{\text L\Omega}^\text W$ in the lift force specifically due to the torque-free rotation, which may be thought of as a Magnus lift component, must first be examined in detail. As Fig. \ref{fig:lift_spin-coef} shows, when normalized by the rotation rate $\Rry^\text{W}$, this change only weakly depends on $\Rey$, especially for small separations. A similar behavior was observed in \cite{2002_Bagchi} in an unbounded shear flow. The rough approximation $C_{\text L\Omega}^\text U/\Rry \approx 0.55$ provided in this reference is in reasonable agreement with present data beyond $\mathcal{O}(1)$-Reynolds numbers, as the solid lines in Fig. \ref{fig:lift_spin-coef} show (the difference is larger at low $\Rey$, as expected from the difference between \eqref{eq:CLWin_tf} and \eqref{eq:CLWin} which predicts $C_{\text L\Omega}^\text U/\Rry \approx 1-\frac{1}{8}\Rry$ for large separations). Based on this finding, one can expect the total lift force acting on a torque-free rotating sphere with $\Rey\gtrsim1$ to be correctly estimated by superposing linearly the force found in the non-rotating case as given by \eqref{eq:CLWmoderRe} and the spin-induced contribution discussed above. This superposition yields
\begin{figure}
	\centerline{\includegraphics[scale=0.65]{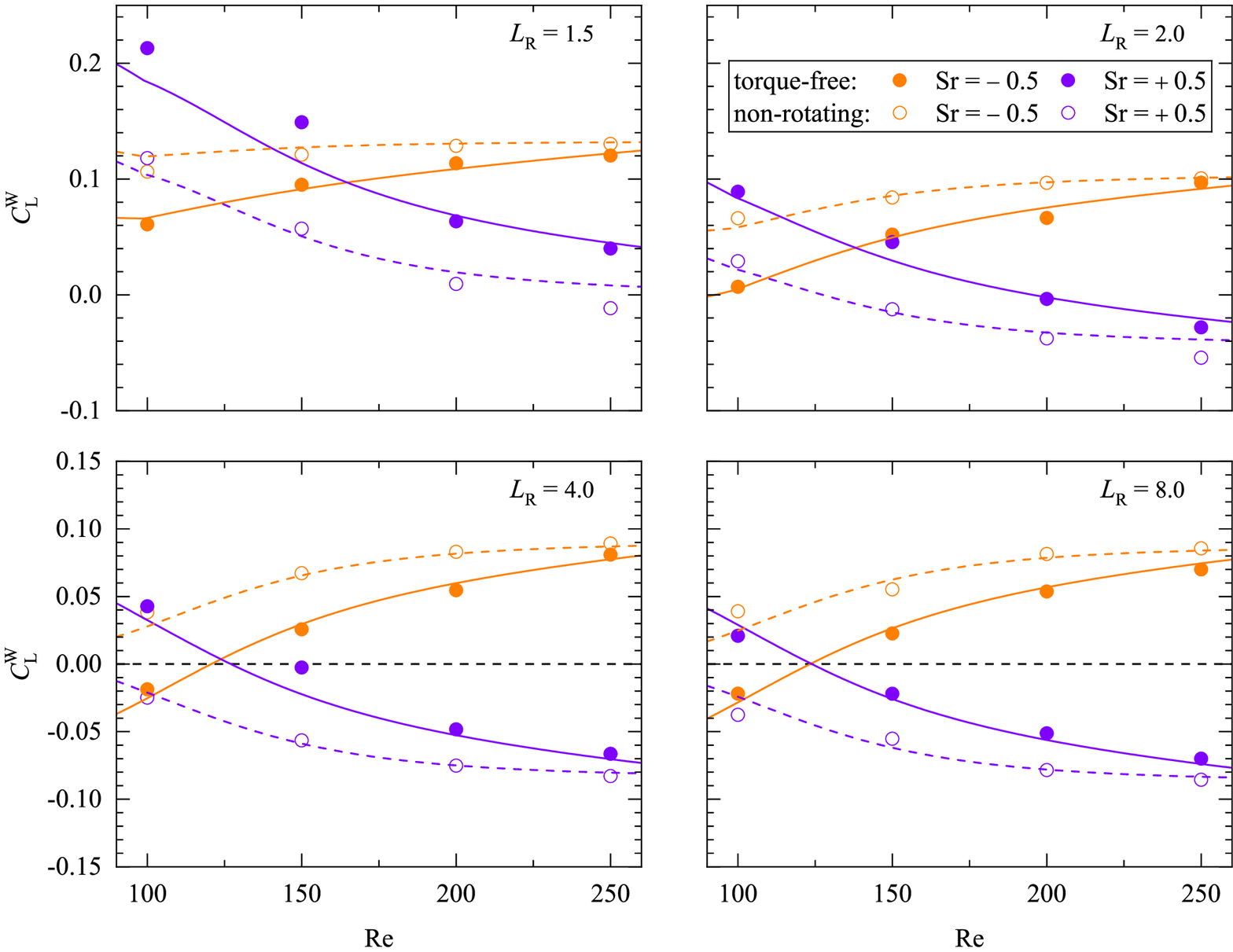}}
	\caption{Influence of the torque-free rotation on the lift force in the high-Reynolds-number regime for $|\Sry|=0.5$. Closed and open symbols refer to data obtained with torque-free and non-rotating spheres, respectively. Solid and dashed lines correspond to the prediction of \eqref{eq:lift_highRe_rot} for the torque-free and non-rotating cases, respectively. The horizontal black dashed line materializes the dividing line $C_\text L^\text W=0$. }
	\label{fig:lift_spin-high-Re}
\end{figure}
\begin{equation}
C_\text L^\text W[\Rey=\mathcal{O}(1-100)] \approx g_\text L C_{\text Lu}^\text W [\Rey=\mathcal{O}(1-100)]+h_\text LC_{\text L\omega}^\text{U}(\Rey\ll1)+0.55\Rry^\text{W}\,,
\label{eq:lift_tf}
\end{equation}
with $\Rry^\text{W}$ as provided in \eqref{eq:spin_rate_w2}. The dashed lines in Fig. \ref{fig:lift_spin} confirm that this linear superposition fits the numerical data well up to $\Rey\approx 100$, even in the low-Reynolds-number range provided the separation is not `too' large.\\
\indent In the high-$\Rey$ regime $\Rey>100$, the total lift force is small, with lift coefficients typically of $\mathcal{O}(0.1)$, \text{i.e.} one order of magnitude smaller than in the low-$\Rey$ regime. 
However, the relative contribution of the change $C_{\text L\Omega}^\text W$ caused by the torque-free rotation in the total lift force remains significant, as Fig. \ref{fig:lift_spin-high-Re} shows. Again, for a given Reynolds number and separation distance, $C_{\text L\Omega}^\text W$ is seen to be larger when $\Sry$ is positive, especially for $L_\text{R}\leq2$. Moreover, the qualitative influence of the sphere rotation is found to depend crucially on the separation distance. Indeed, for $\Rey\geq150$, the torque-free condition is seen to decrease the magnitude of the total lift force irrespective of its sign for $L_\text{R}\geq4$. This is no longer the case at the smallest two separations, for which lift forces corresponding to $\Sry<0$ are still reduced by the rotation while those associated with positive $\Sry$ are enhanced, especially for $L_\text{R}=1.5$. 
To approach the observed behaviors, we again considered that rotation-induced effects combine linearly  with the slip- and shear-induced contributions predicted by \eqref{eq:lift_highRe}, assuming that the empirical expression \eqref{eq:spin_rate_w2} for the rotation rate derived at moderate Reynolds number remains valid up to the upper bound ($\Rey=250$) of the regime considered here. Figure \ref{fig:lift_spin-coef} indicates that the ratio $C_{\text L\Omega}^\text W/\Rry^\text W$ is still close to $0.55$ in this regime, although it seems to rise to slightly larger values ($\approx0.7$) for positive $\Sry$ when the separation becomes small. Keeping this ratio unchanged, we obtain
\begin{equation}
C_\text L^\text W[\Rey=\mathcal{O}(100)] \approx C_{\text Lu}^\text W (\Rey=100)+k_{\text L}C_{\text L\omega}^\text{U}[\Rey=\mathcal{O}(100)]+0.55\Rry^\text{W}\,.
\label{eq:lift_highRe_rot}
\end{equation}
As Fig. \ref{fig:lift_spin-high-Re} shows, this fit reproduces the observed trends well throughout the considered Reynolds number range, although the influence of the torque-free rotation appears to be slightly under-estimated at small separations when $\Sry$ is positive. This successful extension of \eqref{eq:lift_highRe} indicates that effects of slip, shear and torque-free rotation may still be considered to contribute separately to the lift force even for $\mathcal{O}(100)$-Reynolds numbers, provided of course the influence of the nearby wall is properly accounted for in the magnitude of each contribution. 

\section{Summary}\label{sec:6}	
We computed the flow and the hydrodynamic forces acting on a rigid sphere moving along the planar wall bounding a linear shear flow over a wide range of Reynolds number and separation distance, with the sphere either lagging or leading the fluid. We considered both non-rotating and torque-free spheres in order to quantify effects of the rotation induced by the torque-free constraint obeyed by freely moving particles. To reveal the slip-wall and shear-wall interaction mechanisms at stake, we examined several characteristic features of the flow field, especially the spatial distribution of the spanwise and streamwise vorticity disturbances, before focusing on their influence on the drag and lift forces. \\  
\indent When the sphere moves in a fluid at rest, low-Reynolds-number asymptotic solutions indicate an increase of the drag due to the presence of the wall. Additionally, a repulsive transverse force arises, due to the interaction between the wall and the wake resulting from the vorticity generated at the sphere surface by the no-slip condition. For a given separation distance, the magnitude of this repulsive force decreases with the Reynolds number when the wall lies in the outer region of the disturbance, in line with the conclusions of previous studies. At low-but-finite Reynolds number, both the drag increase and the transverse force are proportional to the square of the maximum vorticity at the sphere surface, which increases with the Reynolds number. Present results confirm these predictions, and support the model \eqref{eq:CLuW_moderRe} proposed in \cite{2003_Takemura} for the transverse force up to $\mathcal{O}(100)$-Reynolds numbers, albeit with a slight change in the evaluation of the surface vorticity aimed at accounting for the influence of the nearby wall. At larger Reynolds number, the behavior of the transverse force depends crucially on the separation. For $L_\text{R}\geq4$, this force is nearly zero from Reynolds numbers of some tens up to the critical value $\Rey^{SS}\approx210$ corresponding to the onset of a stationary non-axisymmetric wake. Although the wall is not responsible for this change in the wake structure, it selects the direction of the corresponding lift force, which again tends to repel the sphere into the fluid for $\Rey>\Rey^{SS}$. Up to $\Rey=250$, the magnitude of this force is accurately estimated by the heuristic extension \eqref{eq:CL_Bi} of the theoretical prediction derived from a weakly nonlinear analysis. For smaller separations, the flow past the sphere remains anisotropic whatever the Reynolds number, making the transverse force keep significant values throughout the $\Rey$-range explored numerically. In this situation, the force does not change much beyond $\Rey=100$ when the separation is small ($L_\text{R}=1.5$), while a mixed situation in which the force increases significantly with the Reynolds number in the range $150\lesssim\Rey\lesssim250$ takes place at intermediate separations ($L_\text{R}=2$).

At low Reynolds number, asymptotic predictions with the wall standing in the inner region of the disturbance predict that the shear tends to decrease (increase) the drag when the sphere lags (leads) the fluid, while the reverse holds for the transverse force. For this reason, the latter may switch from positive to negative at a given separation if the sphere leads the fluid and the relative magnitude of the shear is large enough. These predictions are confirmed, both qualitatively and quantitatively, by present numerical results. When the wall stands in the outer region of the disturbance, the semiempirical expressions of \cite{2020_Shi_a} taking into account finite-size effects are found to provide reliable predictions for both the drag variation and the lift force irrespective of the wall position up to $\Rey=2$. Whatever $\Sry$ and $L_\text{R}$, the magnitude of the lift force sharply decreases as the Reynolds number increases in the range $1\lesssim\Rey\lesssim10$. For $L_\text{R}\gtrsim2$, only a weak lift force, with a magnitude close to that found in an unbounded flow, subsists in the moderate-to-high Reynolds number regime $10\lesssim\Rey\lesssim100$. This force keeps significantly larger values at smaller separations, being dominated by the slip effect rather than the influence of the shear in this $\Rey$-range. Numerical results allowed us to obtain the empirical prediction \eqref{eq:CLWmoderRe} for the lift force extending the finite-$\Rey$ prediction \eqref{eq:CWlowRe} up to $\Rey\lesssim100$.\\
\indent At $\mathcal{O}(100)$-Reynolds numbers, considering the unbounded sheared configuration first was found useful to quantify specific effects induced by the wall. Present results confirm the well-established reversal of the shear-induced lift beyond $\Rey\approx50$ \cite{1999_Kurose}. Variations of this `reversed' lift force with $\Rey$ and $\Sry$ agree well with those reported in the literature, as summarized in \cite{2008_Loth}. 
In the same regime, the drag force is found to increase linearly with $\Rey$ and $|\Sry|$ beyond $\Rey\approx150$, leading to a substantial increase ($\approx20\%$) at $\Rey=250$ for $|\Sry|=1$. 
When a nearby wall is involved, the above reversal makes the slip- and shear-related mechanisms contributing to the lift act in an antagonistic (cooperative) manner for positive (negative) $\Sry$, unlike the situation encountered at lower Reynolds numbers. Moreover, these mechanisms interact in a highly nonlinear manner, the shear-induced variation to the lift force observed for a given magnitude of the relative shear rate being significantly larger when $\Sry$ is positive. We could summarize the effect of these complex interactions into the empirical prediction \eqref{eq:lift_highRe}-\eqref{eq:coeffhighRe} which provides an accurate estimate of the near-wall lift force up to $\Rey=250$.\\
\indent 
Only small changes are observed in the flow structure when the sphere rotates in order to satisfy a torque-free condition. The corresponding rotation rate decreases drastically as $\Rey$ increases, similar to the tendency already reported in an unbounded shear flow. However, these small changes subtly modify the shear stress distribution at the sphere surface, hence the torque acting on it. For this reason, they are sufficient to make the variations of the rotation rate with respect to $\Rey$ and $\Sry$ nontrivial in near-wall configurations. First, the slip-induced rotation in a fluid at rest is found to change sign beyond a critical separation-dependent $\mathcal{O}(1)$-Reynolds number. Then, for small enough separations and Reynolds numbers $\gtrsim10$, the rotation rate is influenced by the shear in a very asymmetric manner, depending on the sign of $\Sry$. Indeed, while the rotation is almost identical to its counterpart in an unbounded shear flow when $\Sry$ is negative, it is significantly larger when $\Sry$ is positive, even for Reynolds numbers of $\mathcal{O}(100)$. These findings are summarized in the fit \eqref{eq:spin_rate_w2} which predicts the rotation rate well irrespective of the sign of $\Sry$ and throughout the range of Reynolds number explored in this investigation. Finally, present results show that the spin-induced contribution to the near-wall lift in the torque-free configuration is directly proportional to the rotation rate. Remarkably, the corresponding pre-factor ($\approx0.55$) only weakly varies with the Reynolds number and is similar to that previously determined in an unbounded shear flow \cite{2002_Bagchi}. These findings allow the fits predicting the lift force on a non-rotating sphere to be extended easily to a torque-free sphere in the form \eqref{eq:lift_tf} for moderate Reynolds numbers and \eqref{eq:lift_highRe_rot} for $\Rey\geq100$.\\

\section*{Acknowledgements}			
This work was supported by the Chinese Scholarship Council (CSC). We thank Anna\"ig Pedrono for her intensive help and support with the JADIM code and the grid generator. 

\bibliographystyle{unsrt}
\bibliography{PRF_near-wall_particles}

\begin{thebibliography}{10}

\bibitem{1962_Bretherton}
F.~P. Bretherton.
\newblock The motion of rigid particles in a shear flow at low {R}eynolds
  number.
\newblock {\em J. Fluid Mech.}, 14:284--304, 1962.

\bibitem{1961_Rubinow}
S.~I. Rubinow and J.~B. Keller.
\newblock The transverse force on a spinning sphere moving in a viscous fluid.
\newblock {\em J. Fluid Mech.}, 11:447--459, 1961.

\bibitem{1965_Saffman}
P.~G. Saffman.
\newblock The lift on a small sphere in a slow shear flow.
\newblock {\em J. Fluid Mech.}, 22:385--400, 1965.

\bibitem{1968_Saffman}
P.~G. Saffman.
\newblock Corrigendum to ``{T}he lift on a small sphere in a slow shear flow''.
\newblock {\em J. Fluid Mech.}, 31:624, 1968.

\bibitem{1989_Asmolov}
E.~S. Asmolov.
\newblock Lift force exerted on a spherical particle in a laminar boundary
  layer.
\newblock {\em Fluid Dyn.}, 24:710--714, 1989.

\bibitem{1991_McLaughlin}
J.~B. McLaughlin.
\newblock Inertial migration of a small sphere in linear shear flows.
\newblock {\em J. Fluid Mech.}, 224:261--274, 1991.

\bibitem{1994b_Cherukat}
P.~Cherukat, J.~B. McLaughlin, and A.~L. Graham.
\newblock The inertial lift on a rigid sphere translating in a linear shear
  flow field.
\newblock {\em Int. J. Multiphase Flow}, 20:339--353, 1994.

\bibitem{1999_Cherukat}
P.~Cherukat, J.~B. McLaughlin, and D.~S. Dandy.
\newblock A computational study of the inertial lift on a sphere in a linear
  shear flow field.
\newblock {\em Int. J. Multiphase Flow}, 25:15--33, 1999.

\bibitem{1999_Kurose}
R.~Kurose and S.~Komori.
\newblock Drag and lift forces on a rotating sphere in a linear shear flow.
\newblock {\em J. Fluid Mech.}, 384:183--206, 1999.

\bibitem{2002_Bagchi_2}
P.~Bagchi and S.~Balachandar.
\newblock Shear versus vortex-induced lift force on a rigid sphere at moderate
  re.
\newblock {\em J. Fluid Mech.}, 473:379--388, 2002.

\bibitem{2002_Bagchi}
P.~Bagchi and S.~Balachandar.
\newblock Effect of free rotation on the motion of a solid sphere in linear
  shear flow at moderate {R}e.
\newblock {\em Phys. Fluids}, 14:2719--2737, 2002.

\bibitem{1968_Cox}
R.~G. Cox and H.~Brenner.
\newblock The lateral migration of solid particles in {P}oiseuille flow-{I}
  {T}heory.
\newblock {\em Chem. Eng. Sci.}, 23:147--173, 1968.

\bibitem{1977_Cox}
R.~Cox and S.~Hsu.
\newblock The lateral migration of solid particles in a laminar flow near a
  plane.
\newblock {\em Int. J. Multiphase Flow}, 3:201--222, 1977.

\bibitem{1977_Vasseur}
P.~Vasseur and R.~G. Cox.
\newblock The lateral migration of spherical particles sedimenting in a
  stagnant bounded fluid.
\newblock {\em J. Fluid Mech.}, 80:561--591, 1977.

\bibitem{2003_Takemura}
F.~Takemura and J.~Magnaudet.
\newblock The transverse force on clean and contaminated bubbles rising near a
  vertical wall at moderate {R}eynolds number.
\newblock {\em J. Fluid Mech.}, 495:235--253, 2003.

\bibitem{2005_Zeng}
L.~Zeng, S.~Balachandar, and P.~Fischer.
\newblock Wall-induced forces on a rigid sphere at finite {R}eynolds number.
\newblock {\em J. Fluid Mech.}, 536:1--25, 2005.

\bibitem{2009_Zeng}
L.~Zeng, F.~Najjar, S.~Balachandar, and P.~Fischer.
\newblock Forces on a finite-sized particle located close to a wall in a linear
  shear flow.
\newblock {\em Phys. Fluids}, 21:033302, 2009.

\bibitem{1990_Asmolov}
E.~S. Asmolov.
\newblock Dynamics of a spherical particle in a laminar boundary layer.
\newblock {\em Fluid Dyn.}, 25:886--890, 1990.

\bibitem{1993_McLaughlin}
J.~B. McLaughlin.
\newblock The lift on a small sphere in wall-bounded linear shear flows.
\newblock {\em J. Fluid Mech.}, 246:249--265, 1993.

\bibitem{2009_Takemura_a}
F.~Takemura, J.~Magnaudet, and P.~Dimitrakopoulos.
\newblock Migration and deformation of bubbles rising in a wall-bounded shear
  flow at finite {R}eynolds number.
\newblock {\em J. Fluid Mech.}, 634:463--486, 2009.

\bibitem{2009_Takemura_b}
F.~Takemura and J.~Magnaudet.
\newblock Lateral migration of a small spherical buoyant particle in a
  wall-bounded linear shear flow.
\newblock {\em Phys. Fluids}, 21:083303, 2009.

\bibitem{2003_Magnaudet}
J.~Magnaudet, S.~Takagi, and D.~Legendre.
\newblock Drag, deformation and lateral migration of a bouoyant drop moving
  near a wall.
\newblock {\em J. Fluid Mech.}, 476:115--157, 2003.

\bibitem{1994_Cherukat}
P.~Cherukat and J.~B. McLaughlin.
\newblock The inertial lift on a rigid sphere in a linear shear flow field near
  a flat wall.
\newblock {\em J. Fluid Mech.}, 263:1--18, 1994.

\bibitem{1995_Cherukat}
P.~Cherukat and J.~B. McLaughlin.
\newblock The inertial lift on a rigid sphere in a linear shear flow field near
  a flat wall -- {C}orrigendum.
\newblock {\em J. Fluid Mech.}, 285:407, 1995.

\bibitem{2010_Yahiaoui}
S.~Yahiaoui and F.~Feuillebois.
\newblock Lift on a sphere moving near a wall in a parabolic flow.
\newblock {\em J. Fluid Mech.}, 662:447--474, 2010.

\bibitem{1985_Leighton}
D.~T. Leighton and A.~Acrivos.
\newblock A slow motion of viscous liquid caused by a slowly moving solid
  sphere.
\newblock {\em Z. Angew. Math. Phys.}, 36:174--178, 1965.

\bibitem{1995_Krishnan}
G.~P. Krishnan and D.~T. Leighton.
\newblock Inertial lift on a moving sphere in contact with a plane wall in a
  shear flow.
\newblock {\em Phys. Fluids}, 7:2538--2545, 1995.

\bibitem{2020_Ekanayake_a}
N.~Ekanayake, J.~D. Berry, and D.~J.~E. Harvie.
\newblock Lift and drag forces acting on a particle moving in the presence of
  slip and shear near a wall.
\newblock {\em J. Fluid Mech.}, 915:A103, 2021.

\bibitem{2020_Ekanayake_b}
N.~Ekanayake, J.~D. Berry, A.~D. Stickland, D.~E. Dunstan, I.~L. Muir, S.~K.
  Dower, and D.~J.~E. Harvie.
\newblock Lift and drag forces acting on a particle moving with zero slip in a
  linear shear flow near a wall.
\newblock {\em J. Fluid Mech.}, 904:A6, 2020.

\bibitem{2010_Lee}
H.~Lee and S.~Balachandar.
\newblock Drag and lift forces on a spherical particle moving on a wall in a
  shear flow at finite {R}e.
\newblock {\em J. Fluid Mech.}, 657:89--125, 2010.

\bibitem{2020_Shi_a}
P.~Shi, R.~Rzehak, D.~Lucas, and J.~Magnaudet.
\newblock Hydrodynamic forces on a clean spherical bubble translating in a
  wall-bounded linear shear flow.
\newblock {\em Phys. Rev. Fluids}, 5:073601, 2020.

\bibitem{2018_Asmolov}
E.~S. Asmolov, A.~L. Dubov, T.~V. Nizkaya, J.~Harting, and O.~I. Vinogradova.
\newblock Inertial focusing of finite-size particles in microchannels.
\newblock {\em J. Fluid Mech.}, 840:613--630, 2018.

\bibitem{2008_Loth}
E.~Loth.
\newblock Lift of a solid spherical particle subject to vorticity and/or spin.
\newblock {\em AIAA J.}, 46:801--809, 2008.

\bibitem{2019_Shi}
P.~Shi and R.~Rzehak.
\newblock Lift forces on solid spherical particles in unbounded flows.
\newblock {\em Chem. Eng. Sci.}, 208:115145, 2019.

\bibitem{1967_Goldman_a}
A.~J. Goldman, R.~G. Cox, and H.~Brenner.
\newblock Slow viscous motion of a sphere parallel to a plane wall - {I}.
  {M}otion through a quiescent fluid.
\newblock {\em Chem. Eng. Sci.}, 22:637--651, 1967.

\bibitem{1967_Goldman_b}
A.~J. Goldman, R.~G. Cox, and H.~Brenner.
\newblock Slow viscous motion of a sphere parallel to a plane wall - {II}.
  {C}ouette flow.
\newblock {\em Chem. Eng. Sci.}, 22:653--660, 1967.

\bibitem{Note1}
In (\ref {eq:CLWin}), pre-factors expressed in fractional form were derived
  analytically by Lovalenti in an appendix to \cite {1994_Cherukat}, while
  those expressed in decimal form originate from the fitted value of the force
  computed in the form of a volume integral in the same reference.

\bibitem{1993_Natarajan}
R.~Natarajan and A.~Acrivos.
\newblock The instability of the steady flow past spheres and disks.
\newblock {\em J. Fluid Mech.}, 254:323--344, 1993.

\bibitem{2012_Fabre}
D.~Fabre, J.~Tchoufag, and J.~Magnaudet.
\newblock The steady oblique path of buoyancy-driven disks and spheres.
\newblock {\em J. Fluid Mech.}, 707:24--36, 2012.

\bibitem{2013_Homann}
H.~Homann, J.~Bec, and R.~Grauer.
\newblock Effect of turbulent fluctuations on the drag and lift forces on a
  towed sphere and its boundary layer.
\newblock {\em J. Fluid Mech.}, 721:155--179, 2013.

\bibitem{1965_Happel}
J.~Happel and H.~Brenner.
\newblock {\em Low Reynolds Number Hydrodynamics}.
\newblock Prentice-Hall, 1965.

\bibitem{1995_Magnaudet}
J.~Magnaudet, M.~Rivero, and J.~Fabre.
\newblock Accelerated flows past a rigid sphere or a spherical bubble. {P}art
  1. {S}teady straining flow.
\newblock {\em J. Fluid Mech.}, 284:97--135, 1995.

\bibitem{2003_Legendre}
D.~Legendre, J.~Magnaudet, and G.~Mougin.
\newblock Hydrodynamic interactions between two spherical bubbles rising side
  by side in a viscous liquid.
\newblock {\em J. Fluid Mech.}, 497:133--166, 2003.

\bibitem{2016_Citro}
V.~Citro, J.~Tchoufag, D.~Fabre, F.~Giannetti, and P.~Luchini.
\newblock Linear stability and weakly nonlinear analysis of the flow past
  rotating spheres.
\newblock {\em J. Fluid Mech.}, 807:62--86, 2016.

\bibitem{1997_Legendre}
D.~Legendre and J.~Magnaudet.
\newblock A note on the lift force on a spherical bubble or drop in a
  low-{R}eynolds-number shear flow.
\newblock {\em Phys. Fluids}, 9:3572--3574, 1997.

\bibitem{2002_Takemura}
F.~Takemura, S.~Takagi, J.~Magnaudet, and Y.~Matsumoto.
\newblock Drag and lift forces on a bubble rising near a vertical wall in a
  viscous liquid.
\newblock {\em J. Fluid Mech.}, 461:277--300, 2002.

\bibitem{1964_ONeill}
M.~E. O'Neill.
\newblock A slow motion of viscous liquid caused by a slowly moving solid
  sphere.
\newblock {\em Mathematika}, 11:67--74, 1964.

\end{thebibliography}

\end{document}